\documentclass[%
reprint,
superscriptaddress,
nofootinbib,
amsmath,
amssymb,
aps,
floatfix,
showkeys,
]{revtex4-2}

\usepackage[normalem]{ulem}

\usepackage{fancyhdr}
\usepackage{graphicx,amsfonts,amssymb,amsbsy}
\usepackage{amsmath,amsthm,latexsym}
\usepackage[utf8]{inputenc}  %permitir acentos etc

\DeclareMathOperator\arcsinh{arcsinh}

\usepackage{natbib}
\usepackage[colorlinks]{hyperref}
\hypersetup{linkcolor=magenta,citecolor=blue}

\begin{document}

\title{Cold dense quark matter with phenomenological medium effects: a self-consistent formulation of the quark-mass density-dependent model}

\author{G. Lugones}
\email{german.lugones@ufabc.edu.br}
\affiliation{Universidade Federal do ABC, Centro de Ci\^encias Naturais e Humanas, Avenida dos Estados 5001- Bang\'u, CEP 09210-580, Santo Andr\'e, SP, Brazil.}

\author{A. G. Grunfeld}
\email{grunfeld@tandar.cnea.gov.ar}
\affiliation{CONICET, Godoy Cruz 2290, Buenos Aires, Argentina} 
\affiliation{Departamento de F\'\i sica, Comisi\'on Nacional de Energ\'{\i}a At\'omica, Av. Libertador 8250, (1429) Buenos Aires, Argentina}

\begin{abstract}
We revisit the quark-mass density-dependent model -- a phenomenological equation of state for deconfined quark matter in the high-density low-temperature regime -- and show that thermodynamic inconsistencies that have plagued the model for decades, can be solved if the model is formulated in the canonical ensemble instead of the grand canonical one. Within the new formulation,  the minimum of the energy per baryon occurs at zero pressure, and the Euler's  relation is verified. Adopting a typical mass-formula, we first analyze in detail a simple model with one particle species. We show that a  ``bag'' term that produces quark confinement naturally appears in the pressure (and not in the energy density) due to density dependence of the quark masses. Additionally, the chemical potential gains a new term as in other models with quark repulsive interactions.
Then, we extend the formalism to the astrophysically realistic case of charge-neutral three-flavor quark matter in equilibrium under weak interactions, focusing on two different mass formulae: a flavor dependent and a flavor blind one. For these two models, we derive the equation of state and analyze its behavior for several parameter choices.  We systematically analyze the parameter space and identify the regions corresponding to self-bound 2-flavor and 3-flavor quark matter, hybrid matter and causal behavior.   
\end{abstract}

\maketitle

\section{Introduction}

First principle calculations of deconfined quark matter are not available at present in the high-density low-temperature regime expected in neutron star (NS) interiors. Thus, although perturbative Quantum Chromodynamics (QCD) imposes some constraints on the equation of state (EOS) at NS densities \cite{Annala:2019puf}, most conclusions about quark matter in the NS regime rely on phenomenological models that, inspired by QCD, incorporate in the EOS key properties of quarks such as color confinement, asymptotic freedom and chiral symmetry breaking/restoration. 
In this context, the quark-mass density-dependent model (QMDDM) has been discussed in the literature for decades. 
Its basic idea is that some relevant features of the strong interaction between quarks can be encoded in the appropriate variation of quark masses with the baryon number density $n_B$ and the system as a whole would be described as a non-interacting gas of quasiparticles with density-dependent masses.

The QMDDM was originally introduced by Fowler, Raha, and Weiner to study the properties of quark-matter \cite{Fowler:1981rp} and was then applied by Chakrabarty and coworkers to describe self-bound strange quark matter \cite{Chakrabarty:1989bq,Chakrabarty:1991ui,Chakrabarty:1993db}. In this original version  all  thermodynamic formulae keep exactly the same form as in the constant-mass case. 

Later, the QMDDM was reformulated by Benvenuto and Lugones \cite{Benvenuto:1989kc, Lugones:1995vg} who showed that a new term arises naturally in the pressure when the density dependence of quark masses is taken into account in the thermodynamic derivatives. Indeed, the pressure $P$ was claimed to be given by:
\begin{equation}
P =-\Omega+n_{B} \frac{\partial \Omega}{\partial n_{B}},
\label{eq:new_pressure}
\end{equation}
being $\Omega$ the grand thermodynamic potential. The second term has the same confining role as the bag constant in the MIT bag model, i.e. it makes the pressure to vanish at a finite density.  A new term also appears in the energy density $\epsilon$ via Euler's relation \cite{Benvenuto:1989kc, Lugones:1995vg,Benvenuto:1998tx,Lugones:2002vd}:
\begin{equation}
\epsilon = Ts - P + \sum_i \mu_i n_i ,
\label{eq:Euler}
\end{equation}
where $T$ is the temperature, $s$ the entropy density, $n_i$ the particle number density of the $i-$species, and $\mu_i$ the corresponding chemical potential.
However, that reformulation of the model still contained a thermodynamic inconsistency associated with the energy density formula,  because the minimum of the energy per baryon  does not correspond to the zero pressure point \cite{Benvenuto:1989kc, Lugones:1995vg,Peng:1999gh,Wang:2000dc}, as it should be according to the relation:
\begin{equation}
P= n_B^2 \frac{\partial (\epsilon /n_B)}{\partial n_B}.
\end{equation}

%----------------------------------------------------------------

To circumvent this problem, a different version of the QMDDM was proposed  \cite{Peng:2000ff}. The new term coming from the density dependence of quark masses was retained in the pressure but not in the energy density. To justify this choice it was argued that Euler’s relation should only be used when $P=-\Omega$, but it would not be valid in the QMDDM because in this case the pressure has the extra term shown in Eq. \eqref{eq:new_pressure}. Therefore, according to Ref. \cite{Peng:2000ff}, the energy density in the QMDDM should be obtained from the inverse Legende transformation:
\begin{equation}
\epsilon = \Omega + T s  + \sum_i \mu_i n_i ,
\label{eq:Legendre}
\end{equation}
and not from the Euler's relation shown in Eq. \eqref{eq:Euler}. With that change, it was shown that the energy per baryon minimum occurs now at zero pressure \cite{Peng:2000ff}.

However, as emphasized by Yin and Su \cite{Yin:2008me} the reformulation presented in \cite{Peng:2000ff} is also inconsistent because in their formulae, the variable that are kept constant in the partial derivatives were not treated properly for some quantities. As a consequence, new inconsistencies appear in the pressure, the entropy and the particle number density (see \cite{Yin:2008me} for more details). Moreover,  while the use of Eq. \eqref{eq:Legendre} is correct,  Eq. \eqref{eq:Euler} cannot be taken as invalid, as claimed in \cite{Peng:2000ff}. The reason is that Eq.  \eqref{eq:Euler} is a direct consequence of the fact that the total internal energy $U \equiv V \epsilon$ is an extensive quantity, i.e. it must be  a homogeneous first-order function of the extensive parameters \cite{Callen}:
\begin{equation}
U(\lambda S, \lambda V, \{\lambda N_{i} \} ) = \lambda ~ U (S, V, \{ N_{i} \} ),
\end{equation}
being $\lambda$ an arbitrary constant, $V$ the system's volume, $S = s V$ and $N_i =  n_i V$. Thus,  Eqs. \eqref{eq:Euler} and \eqref{eq:Legendre}  \textit{must both be valid}, and hence the pressure must be always given by $P=- \Omega$. Therefore, there should be some reason that makes Eq. \eqref{eq:new_pressure} wrong.

A possible solution to the above-mentioned problems was proposed by Yin and Su \cite{Yin:2008me} who argued that the medium dependent mass $m^*$, must be taken as a new independent thermodynamic variable.  According to them, $m^*$ must be considered as a constant in the derivatives of the pressure, the entropy and the particle number density (see also \cite{Xia:2014zaa}):
\begin{equation}
\begin{aligned} S &=-\left(\frac{\partial \Omega}{\partial T}\right)_{V, \mu, m^{*}}, & p &=-\left(\frac{\partial \Omega}{\partial V}\right)_{T, \mu, m^{*}} , \\ n &=-\left(\frac{\partial \Omega}{\partial \mu}\right)_{T, V, m^{*}}, & X &=\left(\frac{\partial \Omega}{\partial m^{*}}\right)_{T, V, \mu} , \end{aligned}
\end{equation}
where $X$ is an extensive quantity corresponding to the intensive variable $m^*$. Within this approach, all thermodynamic quantities ($\Omega$, $n_i$, $S$, $p$, $\epsilon$, etc.) take the same functional form as for the standard Fermi ideal gas, but with medium dependent masses. However, negative pressures expected from the physical vacuum do not emerge naturally in this version of QMDDM,  and must be introduced \textit{ad hoc} in the same way as they are in other models (e.g. MIT bag EOS). Thus, one of the most interesting aspects of the first reformulation \cite{Benvenuto:1989kc, Lugones:1995vg} of the QMDDM is lost, which is the natural appearance of confinement directly from the medium dependence of the quasiparticle  masses.

All pathologies of the above versions of the QMDDM arise from the fact that the quark mass is assumed to be a function of $n_B$ but the thermodynamic treatment is made in the frame of the grand canonical ensemble. Such starting point is ill-defined because the grand potential $\Omega$ depends explicitly on $T$,  $V$  and $\mu_i$, but not on $n_B$ or $n_i$, which are derived quantities in the grand canonical representation. 
The natural frame for constructing the QMDDM is the canonical ensemble because  a quark mass, parametrized in terms of $n_B$ or $n_i$, can be introduced consistently in the associated thermodynamic potential (the Helmholtz free energy $F$) which depends naturally on $T$, $V$ and  $n_i$. 
In the present paper we will reformulate the QMDDM in the canonical ensemble and show that a self-consistent EOS can be derived while maintaining key properties of quark matter such as confinement, asymptotic freedom and chiral symmetry breaking/restoration. 

Thermodynamic inconsistencies arising from medium dependent single particle characteristics are known in the theory of nuclear matter since long ago.   For example, in the mean-field approximation, a quasinucleon dispersion relation of the form $E^{*}(k, T, \mu)=\left(k^{2}+M^{2}\right)^{1 / 2}+U$ with an effective nucleon mass $M(T, \mu)$ and potential energy $U(T, \mu)$ appears due to the presence of the interaction between nucleons and the scalar and vector fields. These fields contribute to the effective Hamiltonian of the system and, therefore, produce additional field terms in the pressure and the energy density to restore the thermodynamical consistency of the model \cite{Gorenstein:1995vm}.
The same occurs when medium effects are included in models of cold degenerate quark matter making use of effective quark masses derived from the zero momentum limit of the dispersion relations following from an effective quark propagator obtained from resumming one-loop self-energy diagrams in the hard dense loop approximation.  Due to the $\mu$-dependence of the effective mass, a counterterm is added to the effective Hamiltonian in order to maintain thermodynamic self-consistency.  From this Hamiltonian the energy density  and pressure at zero temperature of a Fermi gas of free quasiparticles are given by expressions that contain an extra term  \cite{Schertler:1996tq,Wen:2010zz}.
Notice that in the above ``quasiparticle models'' the effective masses are parametrized in terms of $\mu_i$ and/or $T$ and the addition of an extra term is sufficient to solve thermodynamic inconsistencies within the grand canonical ensemble. However, the same procedure has been tried to solve the QMDDM without success as explained above.

This paper is organized as follows. In Sec. \ref{sec:2} we review some well-known results for degenerate Fermi gases with medium independent particle masses in order to set the notation and introduce some functions that are used along the paper.  In Sec. \ref{sec:3} we reformulate the QMDDM in the canonical ensemble for the simple case of a system with only one flavor. We show that thermodynamic inconsistencies do not arise in the model; for example, the minimum of the energy per baryon occurs at zero pressure and the Euler's relation is verified. It is also shown that a  ``bag'' term that produces quark confinement naturally appears in the pressure (and not in the energy density) due to density dependence of the quark masses. Additionally, the chemical potential gains a new term as in other models with quark repulsive interactions.
In Sec. \ref{sec:4} we extend the QMDDM to the more realistic case of charge-neutral three-flavor quark matter in equilibrium under weak interactions. We introduce two different mass formulae -- a flavor dependent and a flavor blind one -- and derive analytic formulae for the EOS. 
In Sec. \ref{sec:5} we solve numerically the EOS for several parameter choices.  We systematically analyze the parameter space and identify the regions corresponding to self-bound 2-flavor and 3-flavor quark matter, hybrid matter and causal behavior.   
In Sec. \ref{sec:6} we summarize our main results and discuss some of the consequences of the model.

%------------------------------------------------------------------------
\section{Degenerate free Fermi gas in different statistical ensembles}
\label{sec:2}
%------------------------------------------------------------------------

In order to develop a thermodynamically consistent formulation of the QMDDM one must work within the Helmholtz representation associated with the canonical ensemble. We review here some well-known results for degenerate Fermi gases with medium-independent particle masses. We begin establishing the notation and introducing some functions that will be used along the paper. For simplicity, we start by considering a system with only one particle species.

%------------------------------------------------------------------------
\subsection{Degenerate free Fermi gas in the grand canonical ensemble}
%------------------------------------------------------------------------

The thermodynamic behavior of a free Fermi gas  is usually described within the grand canonical ensemble. 
From the grand canonical partition function one obtains the grand thermodynamic potential $\Omega(T, V, \mu)$, where $T$ is the temperature, $V$ is the system's volume, and $\mu$ is the chemical potential.  
For relativistic fermions of mass $m$ at $T=0$, $\Omega(V, \mu)$ reads \cite{Shapiro:1983du}:
\begin{equation}
\Omega(V,\mu)  =  -\frac{1}{3} \frac{g V}{2 \pi^{2}} \int_{0}^{k_F}  \frac{\partial E(k)}{\partial k}   k^3 d k = - g V m^4 \phi(x),
\label{eq:Omega}
\end{equation}
where $g$ is the particle's degeneracy and  
\begin{equation}
\phi(x)  = \frac{1}{48 \pi^2}\left[x \sqrt{x^{2}+1}\left(2 x^{2}-3\right)+3 \arcsinh(x)\right],
\label{eq:phi}
\end{equation}
being 
\begin{equation}
x(\mu) \equiv \frac{k_F}{m} = \sqrt{\frac{\mu^2}{m^2} -1} ,
\label{eq:x_grand_canonical}
\end{equation}
with $k_F$ the Fermi momentum.

All thermodynamic quantities can be derived from $\Omega(V, \mu)$. 
The particle number density $n$ is given by:
\begin{align} 
n(\mu) \equiv \frac{N(V, \mu)}{V} =  - \frac{1}{V}\frac{\partial \Omega}{\partial \mu}\bigg|_{V}=  \frac{g m^3}{6 \pi^{2}} x^3, \label{eq:N_grand_canonical}
\end{align}
and the pressure $p$ is:
\begin{align} 
p(\mu)  =  - \frac{\partial \Omega}{\partial V}\bigg|_{\mu} =  g m^4 \phi(x ) .
\label{eq:p_grand_canonical}
\end{align}
Finally, the internal energy $U$ can be obtained from the Legendre's  transform $\Omega  \equiv   U - T  S  -\mu N$. Dividing $U$ by the system's volume, one obtains the energy density $\epsilon \equiv U/V$
\begin{equation}
\epsilon(\mu) = g m^4 \chi(x),
\label{eq:U_grand_canonical}
\end{equation}
where  
\begin{equation}
\chi(x)  = \frac{1}{16 \pi^2}\left[ x \sqrt{x^2 + 1} (2 x^2 + 1) -  \arcsinh(x) \right] .
\label{eq:chi}
\end{equation}
%

%----------------------------------------------------------------
\subsection{Degenerate free Fermi gas in the canonical ensemble}
%----------------------------------------------------------------

In the canonical ensemble all thermodynamic quantities can be derived once the Helmholtz free energy $F(T, V,N)$ is specified. $F(T, V,N)$ is defined as the Legendre transform of the internal energy $U(S, V, N)$ with respect to the entropy $S$, i.e.  $F  \equiv U - T  S$. Thus, at zero temperature we obtain from Eq. \eqref{eq:U_grand_canonical}:
\begin{equation}
F(V, N)  = \epsilon V   = g V m^4 \chi(x)    \qquad (\text{for } T=0) ,
\label{eq:Helmholtz_1flavor}
\end{equation}
where $\chi$ is given by Eq. \eqref{eq:chi} and $x$ is now expressed in terms of the particle number density $n = N/V$ (using Eq. \eqref{eq:N_grand_canonical}):
\begin{equation}
x(n)  = \frac{1}{m} \left( \frac{6 \pi^2 n}{g} \right)^{1/3} .
\label{eq:x_canonical}
\end{equation}

Since at $T=0$ we have $U=F$, the energy density $\epsilon = U/V = F/V$ reads:
\begin{equation}
\epsilon(n) =  g m^4 \chi(x)  .  
\label{eq:Legendre_energy_density}
\end{equation}

The pressure can be obtained from the standard thermodynamic derivative of $F(V, N)$:
\begin{eqnarray}
p(n) &=& - \frac{\partial F}{\partial V}\bigg|_{N} =  - \frac{\partial (\epsilon V/N)}{\partial (V/N)}\bigg|_{N} =  n^2 \frac{\partial (\epsilon /n)}{\partial n}  \label{eq:p_definition}\\
 &=&  g m^4 \left( - \chi + n \chi^{\prime} \frac{\partial x}{\partial n} \right) .
\end{eqnarray}
Replacing $\chi$ from Eq. \eqref{eq:chi}, ${\partial x}/{\partial n} = x/(3n)$, and:
\begin{eqnarray}
\chi^{\prime}(x) \equiv \frac{d \chi}{dx}  =    \frac{x^2  \sqrt{x^2 + 1}    }{  2 \pi^2 }  ,
\label{eq:chi_prime}
\end{eqnarray}
one finds:
\begin{eqnarray}
p(n) &=&   g m^4 \phi(x ) ,
\label{eq:pressure_canonical} 
\end{eqnarray}
where $\phi(x)$ was already defined in Eq. \eqref{eq:phi} and $x$ is given by Eq. \eqref{eq:x_canonical}.
Notice that, although Eq. \eqref{eq:pressure_canonical} coincides formally with the grand canonical result  given in Eq. \eqref{eq:p_grand_canonical}, the variable $x$ in Eq. \eqref{eq:pressure_canonical} is now regarded as a function of $n$ and not as a function of $\mu$.

The chemical potential is given by:
\begin{eqnarray}
\mu  &=&  \frac{\partial F}{\partial N}\bigg|_{V} = \frac{\partial (F/V)}{\partial (N/V)}\bigg|_{V} = \frac{\partial \epsilon}{\partial n}    \label{eq:mu_canonical1} \\
&= &    g m^4 \chi^{\prime} \frac{\partial x}{\partial n}  =   m \sqrt{x^2 + 1}  .
\label{eq:mu_canonical2}
\end{eqnarray} 
Again, the above result coincides with Eq. \eqref{eq:x_grand_canonical}, but $x$ is  given now by Eq. \eqref{eq:x_canonical}.

In summary, at $T=0$, all thermodynamic quantities are functions of the particle number density $n$. The complete EOS reads:
\begin{eqnarray}
x(n) &=& \frac{1}{m} \left( \frac{6 \pi^2 n}{g} \right)^{1/3}, \\
\epsilon(n) &=& g m^4 \chi(x), \\
p(n)  &=&  g m^4 \phi(x),  \\
\mu(n) &=&   m \sqrt{x^2 + 1} .
\end{eqnarray}

%--------------------------------------------------------------
\section{Revisiting the QMDDM for a one-component cold gas}
\label{sec:3}
%--------------------------------------------------------------

%%%%%   FIGURE 1
\begin{figure}[tb]
\centering
\includegraphics[width=\columnwidth]{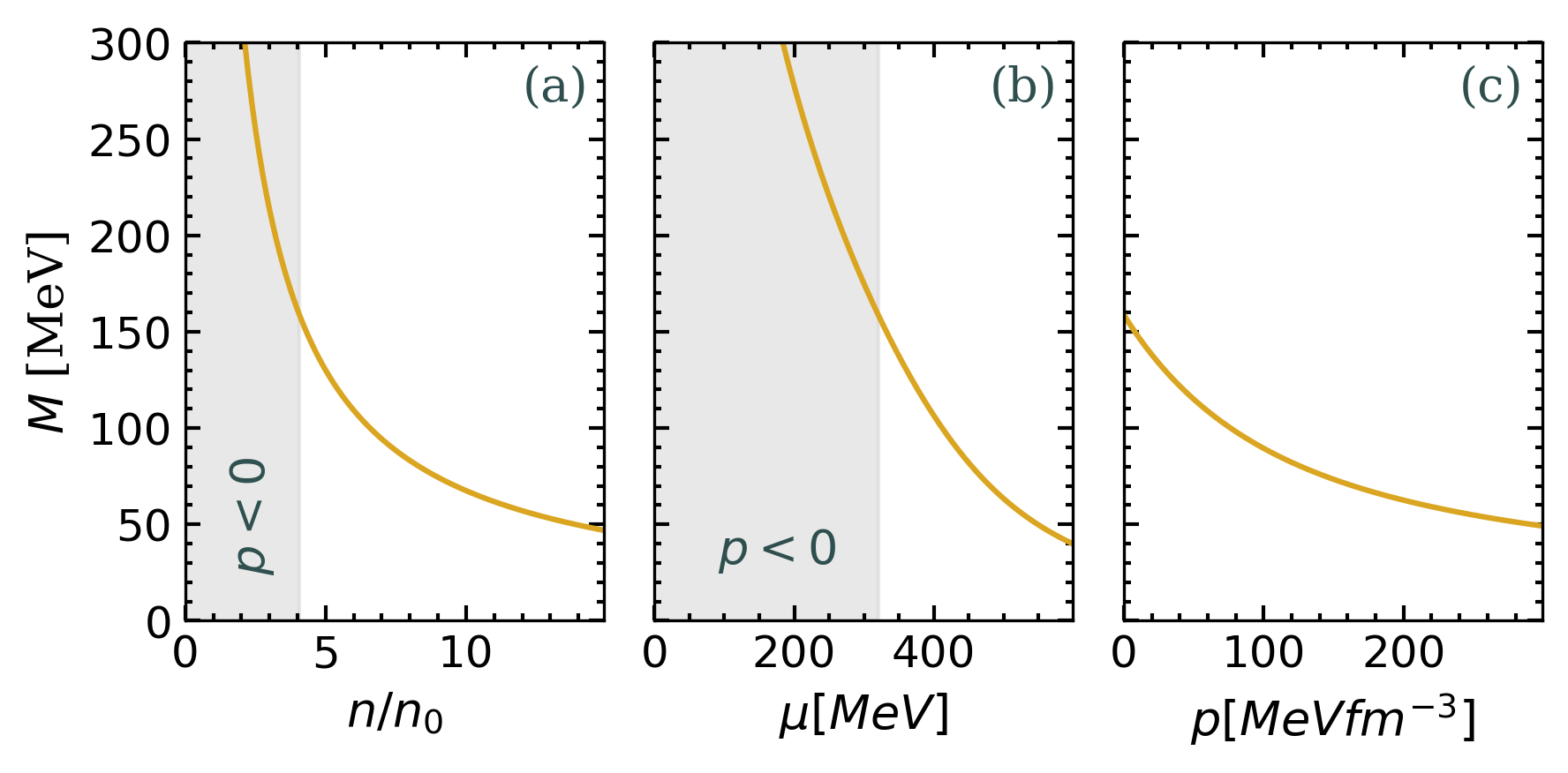}
\caption{Effective quasiparticle mass $M$ as a function of: (a) the particle number density $n$ in units of the nuclear saturation density $n_0$, (b) the chemical potential, and (c) the pressure.  For these plots we adopted $m= 5~ \mathrm{MeV}$, $a=3$ and $C=100$ (units of $a$ and $C$ are such that $n$ is in $\mathrm{fm}^{-3}$ and  $M$ in $\mathrm{MeV}$). Notice that according to the mass formula, $M$ diverges as $n \rightarrow 0$. However, in practice, $M$ has its maximum finite value at zero pressure  and tends to the current mass $m$ at asymptotically large pressures. The gray shaded zones indicate the unphysical regions of negative pressure.  }
\label{fig:1}
\end{figure}

In this section we will reformulate the QMDDM in the simplest case of a one-species system. The central idea of this model is that important aspects of the strong interaction among quarks in the high-density low-temperature regime can be described by treating the system as a free gas of quasiparticles with density-dependent effective masses. Clearly, it is not possible to mimic \textit{all} QCD features in the effective masses and more sophisticated versions of the model including interaction terms would be required in order to extend the applicability range of the model.  

Although this idea is widely known in physics,  the implementation of quasiparticle mass formulae depending on the baryon number density $n_B$ has been subject to long standing problems of thermodynamic consistency (see \cite{Benvenuto:1989kc,Lugones:1995vg,Peng:1999gh,Peng:2000ff,Wang:2000dc,Yin:2008me,Xia:2014zaa,Wen:2010zz} and references therein). The source of all these issues is that density-dependent functions were always introduced within the grand canonical ensemble where $\Omega$ explicitly depends on $T$, $V$ and $\mu$, being the particle number density a derived quantity. Then, it is ambiguous how to deal with a density-dependent mass when thermodynamic derivatives are performed. As we will show in this section, such ambiguity disappears when density-dependent masses are introduced within the canonical ensemble where $F$ depends on $T$, $V$ and the number of particles $N$.

%--------------------------------------------------------------
\subsection{Phenomenological quasiparticle mass}

For the effective quasiparticle mass we will adopt the following ansatz: 
\begin{eqnarray}
M(n) = m + \frac{C}{n^{a/3}} , 
\label{eq:mass_formula}
\end{eqnarray}
where $m$ is the current quark mass, $a$ and $C$ are positive constant free parameters,  and $n = N /V$ is the particle number density. In the next section, we will present a model for 3-flavor quark matter with the mass parametrized in terms of the particle number densities $n_i$ ($i=u, d, s$) or the baryon number density $n_B$.  

The ansatz of Eq. \eqref{eq:mass_formula} has been used in almost  all previous versions of the QMDDM as well as in other phenomenological models \cite{Kaltenborn:2017hus}. 
The effective mass diverges for densities approaching zero (see Fig. \ref{fig:1}), thus mimicking confinement by removing the occurrence of quasiparticles.  At zero pressure $M$ has its largest physical value as seen in Fig. \ref{fig:1}.
At asymptotically large densities the effective mass tends to the current quark mass resembling the restoration of the chiral symmetry in a phenomenological way when $m=0$.

Finally, since we will need the first and second derivatives of $M(n)$ to obtain some thermodynamic quantities, we present them here:
\begin{equation}
\frac{\partial M}{\partial n} =  -\frac{C}{3} \frac{a}{n^{a / 3+1}}, 
\label{eq:mass_derivative}
\end{equation}
\begin{equation}
\frac{\partial^2 M}{\partial n^2} = \frac{C}{9} \frac{a(a+3)}{n^{a / 3+2}} .
\label{eq:2nd_mass_derivative}
\end{equation}
Notice that ${\partial M}/{\partial n} $ is always negative for $n>0$.

%--------------------------------------------------------------
\subsection{QMDDM in the canonical ensemble}

In the following we reformulate the QMDDM in the canonical ensemble assuming that the system can be described as Fermi gas of \textit{free} quasiparticles of mass $M(n)$. Thus, our starting point is the Helmholtz free energy given in Eq. \eqref{eq:Helmholtz_1flavor}, but using $M(n)$ instead of $m$:
\begin{equation}
F(V,N) =  g V M^4(n) \chi(x)  .
\label{eq:Helmholtz_free_energy_2}
\end{equation}
Notice that  $x$ is given  by the same expression of Eq. \eqref{eq:x_canonical} but now with a density dependent mass: 
\begin{equation}
x(n)  = \frac{1}{M(n)} \left( \frac{6 \pi^2 n}{g} \right)^{1/3} .
\label{eq:x_1flavor}
\end{equation}

From Eq. \eqref{eq:Helmholtz_free_energy_2} we can derive all the thermodynamic quantities that describe the macroscopic behavior of the system. In particular, the energy density at $T=0$ is given by  $\epsilon = F/V$ and reads: 
\begin{equation}
\epsilon(n) =  g M^4 \chi(x)  .
\label{eq:energy_density}
\end{equation}
The pressure, chemical potential and speed of sound will be derived in the following subsections.

%%%%%   FIGURE 2
\begin{figure}[tb]
\centering
\includegraphics[width=\columnwidth]{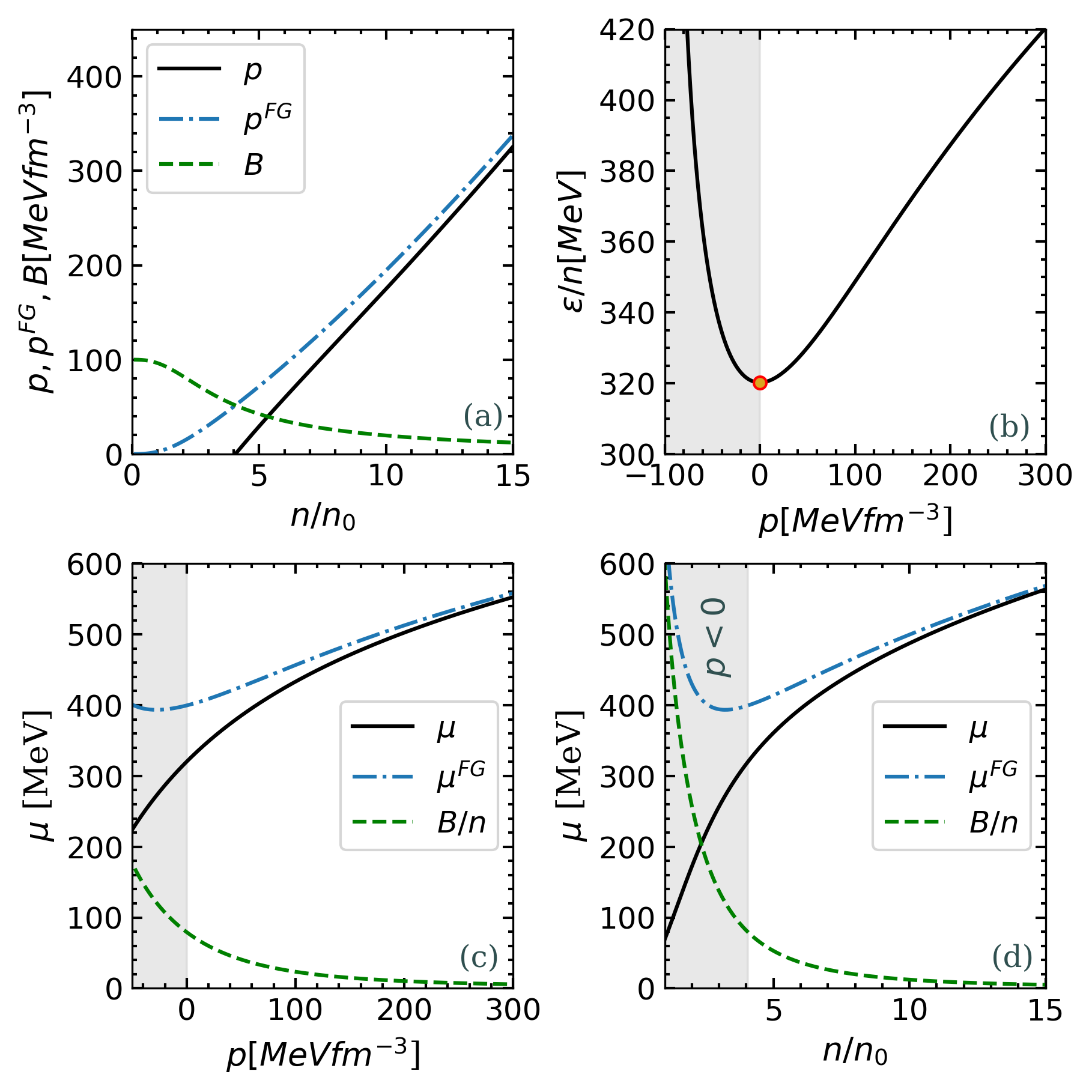}
\caption{EOS for a one component gas: (a) total pressure $p$ (and its contributions  $p^{FG}$ and $B$)  as a function of the particle number density $n$ in units of the nuclear saturation density $n_0$; (b) energy per particle $\epsilon/n$ as a function of $p$; (c) chemical potential $\mu$ and its contributions $\mu^{FG}$ and $B/n$ as a function of $p$; 
(d) chemical potential as a function of $n/n_0$. Notice that $\epsilon/n$ has a minimum at $p=0$, as required by thermodynamic consistency (see red circle in panel b). The gray shaded zones indicate the unphysical regions of negative pressure.}
\label{fig:2}
\end{figure}

%--------------------------------------------------------------
\subsection{Pressure}

To obtain the pressure we use the standard expression given in Eq. \eqref{eq:p_definition}
\begin{eqnarray}
p(n) =  n^2 \frac{\partial (\epsilon /n)}{\partial n}, 
\end{eqnarray}
from which we find: 
\begin{equation}
\begin{aligned}
p(n) =   - g M^4 \chi + g n 4 M^{3} \frac{\partial M}{\partial n} \chi + g n M^4 \chi^{\prime} \frac{\partial x}{\partial n} ,
\end{aligned}
\label{eq:p_1flavor}
\end{equation}
where $\chi^{\prime}$ is given by Eq. \eqref{eq:chi_prime}, and:
\begin{eqnarray}
\frac{\partial x}{\partial n} &=&  \frac{x}{3 n}   -\frac{x}{M} \frac{\partial M}{\partial n} .
\label{eq:dxdn} 
\end{eqnarray}
Replacing Eq. \eqref{eq:dxdn} into Eq. \eqref{eq:p_1flavor} and rearranging terms we obtain: 
\begin{equation}
p(n) = p^{\mathrm{FG}}(n) - B(n)
\label{p_1flavor_final}
\end{equation}
being
\begin{eqnarray}
p^{\mathrm{FG}}(n) & \equiv &    g M^4 \phi(x), \label{pressure_FG_1flavor} \\ 
B(n) & \equiv& - g M^3 n  \frac{\partial M}{\partial n} \beta(x) > 0, \label{eq:bag_1flavor}
\end{eqnarray}
where $\phi(x) = -\chi + x \chi^{\prime}/3$ is given by Eq. \eqref{eq:phi}, and the new function $\beta(x) \equiv 4 \chi - x \chi^{\prime}$ is given by:
\begin{equation}
\beta(x) \equiv  \frac{1}{4 \pi^2}\left[  x \sqrt{x^{2}+1}- \arcsinh(x) \right].
\label{eq:beta}
\end{equation}

The total pressure in Eq. \eqref{p_1flavor_final} has two terms. The first one, given in  Eq. \eqref{pressure_FG_1flavor}, is formally identical to the one shown in Eq. \eqref{eq:pressure_canonical} but with a density dependent mass. Thus, we interpret $p^{\mathrm{FG}}$ as the ``free gas" contribution to the pressure. The new term $B$, given in Eq. \eqref{eq:bag_1flavor}, comes from the density dependence of the quasiparticle's mass $M$. Since $\beta(x)$ is always positive for $x>0$, and $\partial M/\partial n <0$,  we obtain $B(n)>0$.  Therefore, the pressure contains now a new term that is always negative, which is interpreted as a  ``bag constant" that confines the particles. The bag constant vanishes at asymptotically large densities and the system becomes a free Fermi gas of particles with  $M = m$.  The total pressure and its contributions  $p^{FG}$ and $B$ are shown in Fig. \ref{fig:2}$a$. Notice that, due to the effect of $B$, the total pressure vanishes at a finite density.  

An important feature of the model is that confinement naturally arises as a consequence of the density dependence  of the particle's masses.  In contrast with previous versions of the QMDDM, this feature doesn't spoil the thermodynamic consistency of the model. For example,  the energy per particle as a function of the total pressure has its minimum located at $p=0$ (see Fig. \ref{fig:2}$b$).

%--------------------------------------------------------------
\subsection{Chemical potential}
\label{sec:mu_1fl}

The chemical potential is given by (cf. Eqs. \eqref{eq:mu_canonical1} and \eqref{eq:energy_density}):

\begin{eqnarray}
\mu(n)  &=&  \frac{\partial \epsilon}{\partial n} = 
 g M^{4} \chi^{\prime} \frac{\partial x}{\partial n}   + g 4 M^{3} \frac{\partial M}{\partial n} \chi  .
\label{eq:mu_1flavor}
\end{eqnarray}
Replacing  $\chi^{\prime}$ by Eq. \eqref{eq:chi_prime},  $\partial x / \partial n$ by Eq. \eqref{eq:dxdn}, and rearranging terms we obtain: 
\begin{equation}
\mu(n)  =  \mu^{\mathrm{FG}}(n) - \frac{B(n)}{n}
\label{eq:mu_QMDDM_1fl}
\end{equation}
where  
\begin{equation}
\mu^{\mathrm{FG}}(n)  = M \sqrt{x^{2}+1} 
\label{eq:mu_42}
\end{equation}
and the bag constant was already given in Eq. \eqref{eq:bag_1flavor}.

The chemical potential in Eq. \eqref{eq:mu_QMDDM_1fl} has two terms. The first one, $\mu^{\mathrm{FG}}$, has the same form as for a free Fermi gas but with $M$ instead of $m$ (cf. Eq. \eqref{eq:mu_canonical2}).
This contribution comes from the first term of Eq. \eqref{eq:mu_1flavor} which depends on $\partial x/\partial n$; i.e. it  takes into account the variation of the Fermi's momentum $x$ with the density. Usually, the Fermi's momentum increases with $n^{1/3}$ but in the QMDDM it has an extra density dependence embedded in the quasiparticle's mass, as seen in Eq. \eqref{eq:x_1flavor}. Writing Eq. \eqref{eq:mu_42}  as a function of the density, one finds $\mu^{\mathrm{FG}}(n)  = M(n) \sqrt{x(n)^{2}+1} = [ \left({6 \pi^{2} n}/{g}\right)^{2/3}+m^{2}+2 m C n^{-a / 3} + C^{2} n^{-2 a/3} ]^{1/2}$. The first two terms inside the square root are the same as for a Fermi gas with a medium independent quark mass. The third and fourth terms \textit{increase} the chemical potential with respect to the standard case,   taking into account that the Fermi surface changes differently when medium effects are included in $M$. 

The second term, $B/n$,  of  Eq. \eqref{eq:mu_QMDDM_1fl}  comes from the second term of  Eq. \eqref{eq:mu_1flavor} which depends on $\partial M/\partial n$.  Contrary to the first term, it is negative; i.e. it \textit{decreases} the chemical potential. It takes into account that the effect of adding a new particle (at fixed $V$) is not simply to put it at the top of the Fermi's sea. There is an extra effect related to the mass shrink of all particles. As seen in Figs. \ref{fig:2}$cd$, the term $B/n$ vanishes at asymptotically large densities.

%---------------------------------------------------------
\subsection{Speed of sound}

%%%%%   FIGURE 3
\begin{figure}[tb]
\centering
\includegraphics[width=\columnwidth]{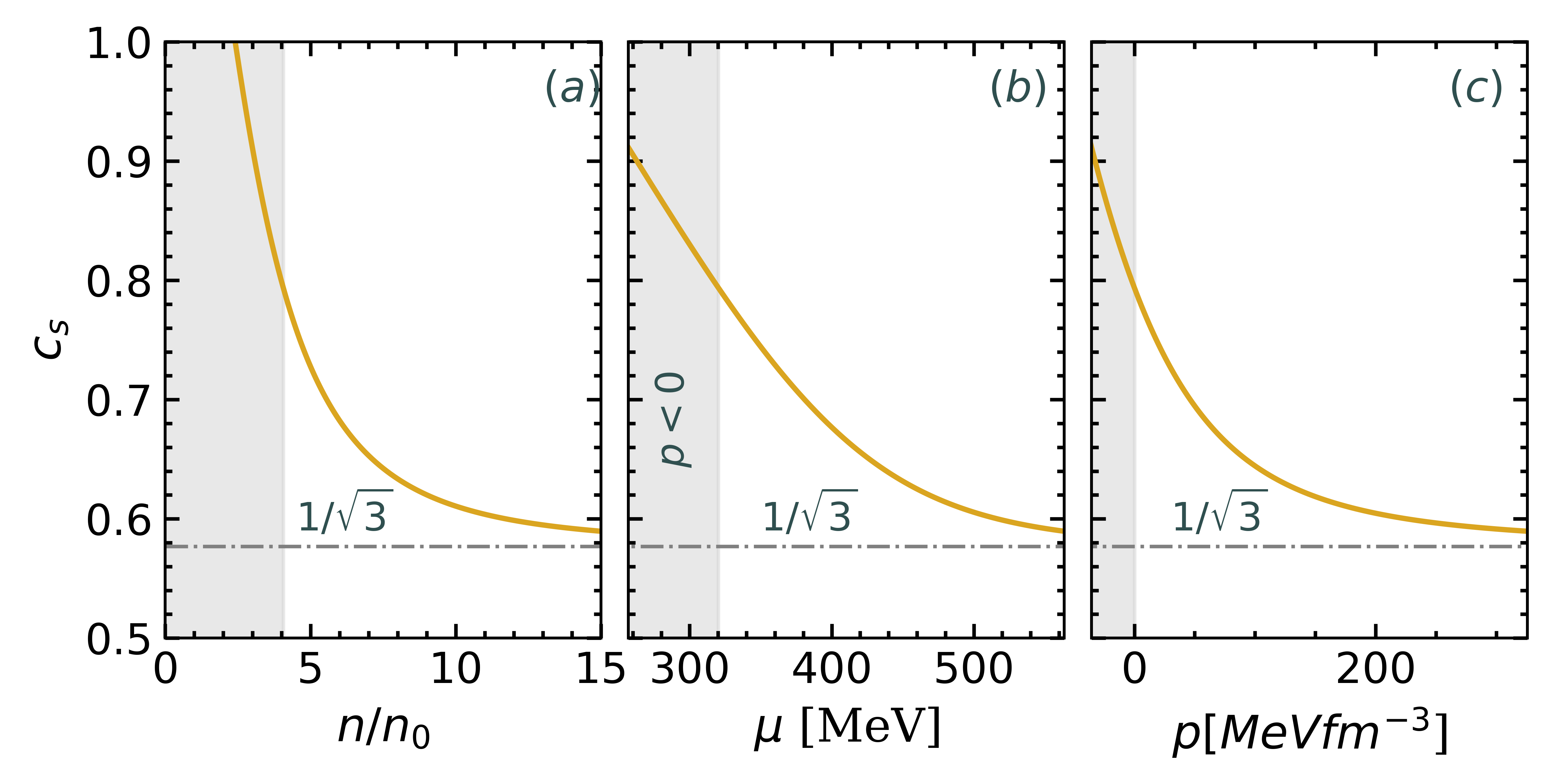}
\caption{Speed of sound for the same choices  of $a$ and $C$ of Fig. \ref{fig:1}. At asymptotically large densities $c_s$ tends to the conformal limit $1/\sqrt{3}$. }
\label{fig:3}
\end{figure}

The sound speed is defined as:
\begin{eqnarray}
c_s^2 = \frac{\partial p}{\partial \epsilon} \bigg|_{S}  .
\end{eqnarray}
At zero temperature the system has $S=0$; therefore, we have:
\begin{eqnarray}
c_s^2  & = &  \frac{\partial p(n)}{\partial \epsilon(n)} 
=  \frac{\partial p / \partial n}{\partial \epsilon/  \partial n} =  
\frac{1}{\mu} \left[ \frac{\partial p^{FG}}{\partial n}  - \frac{\partial B}{\partial n}  \right] ,
\end{eqnarray}
where:
\begin{eqnarray}
\frac{\partial p^{FG} }{\partial n}  &=& g M^3 \frac{\partial M}{\partial n} (4 \phi - \phi^{\prime} x) + \frac{g M^4}{3 n}  \phi^{\prime} x , \\
\frac{\partial B}{\partial n}   &=& - g M^2 n \left(  \frac{\partial M}{\partial n}  \right)^2 [3 \beta - \beta^{\prime} x]  \nonumber \\
    && - \frac{g M^3}{3}   \frac{\partial M}{\partial n}  [3 \beta + \beta^{\prime} x]  - g M^3 n \frac{\partial^2 M}{\partial n^2} \beta ,    
\end{eqnarray}
being:
\begin{eqnarray}
\phi^{\prime}(x)  & = &  \frac{x^4}{  6 \pi^2 \sqrt{x^2 + 1}   }     .
\end{eqnarray}

The speed of sound is shown Fig. \ref{fig:3} for the same choices  of the parameters  $a$ and $C$ used in  Fig. \ref{fig:1}. At asymptotically large densities $c_s$ tends to the conformal limit $1/\sqrt{3}$ and at low densities it increases significantly (we adopt $c=1$, being $c$ the speed of light). 

%-------------------------------------
\section{Generalization to three-flavor quark matter: formalism}
\label{sec:4}
%-------------------------------------

For astrophysical applications, the most relevant state of quark matter is an electrically neutral mixture of $u$, $d$ and $s$ quarks together with a small fraction of electrons $e$, all in chemical equilibrium under weak interactions. The one-flavor formula of the previous section can be extended to three flavors in different ways. In Sec. \ref{sec:4a} we discuss some theoretical motivations for the flavor blind and the flavor dependent  mass formulae that will be used in Secs. \ref{sec:4b} and \ref{sec:4c}. For convenience, we summarize both EOS in Sec. \ref{sec:summary_of_EOS}.  The chemical equilibrium and charge neutrality conditions that will be used in the calculations of the following sections are discussed in Sec. \ref{sec:4e}.

%-------------------------------------
\subsection{Quasiparticle mass formulae for 3-flavor quark matter}
\label{sec:4a}
%-------------------------------------

In the context of the QMDDM, the mass formula for each flavor has been written in previous works as a function of the baryon number density $n_B$. This choice is based on an analysis of a schematic form of the QCD Hamiltonian density, which is written as \cite{Peng:1999gh}:
\begin{equation}
H_{\mathrm{QCD}}=H_{\mathrm{kin}}+\sum_{i} m_{i} \bar{q}_i q_i + H_{\mathrm{int}},
\end{equation}
where $H_{\mathrm{kin}}$ is the kinetic term, $m_{i}$ is the quark current mass of flavor $i$, $H_{\mathrm{int}}$ is the interaction contribution, and the summation goes over all flavors considered. The original Hamiltonian density can be replaced by an equivalent one of the form
\begin{equation}
H_{\mathrm{eq}}= H_{\mathrm{kin}} + \sum_{i} M_{i} \bar{q}_i q_i ,
\end{equation}
where $M_{i}$ is an equivalent mass to be determined. Requiring that  $H_{\mathrm{eq}}$ and $H_{\mathrm{QCD}}$ have the same eigenenergies for any eigenstate $|\Psi\rangle$, and assuming that $(M_i -m_{i})$ is the same for all flavors, $M_{i}$ turns out to be (see \cite{Peng:1999gh} for further details):
\begin{align}
M_{i} &=m_{i}+\frac{    \left\langle n_{B}\left|H_{\mathrm{int}}\right| n_{B}\right\rangle    -   \left\langle 0\left|H_{\mathrm{int}}\right| 0\right\rangle   }   {\sum_{q}\left[    \left\langle n_{B}|\bar{q}_i q_i| n_{B}\right\rangle  -  \langle 0|\bar{q}_i q_i| 0\rangle     \right]} \label{eq:eqivalent_mass} \\  & \equiv m_{i}  +m_{\mathrm{int}} , 
\end{align}
where  $\left|n_{B}\right\rangle$ is a state  with baryon number density $n_{B}$ and  $|0\rangle$ is the vacuum state. The first term of Eq. \eqref{eq:eqivalent_mass}  is the original current mass, while $m_{\mathrm{int}}$ is a contribution arising from interactions. Since $m_{\mathrm{int}}$ is the ratio of the total interacting part of the energy density and the total relative quark condensate, it is flavor independent and density dependent. 

In Eq. (44) of \cite{Wen:2005uf} it is shown that the numerator of Eq. \eqref{eq:eqivalent_mass} is given by $3 n_{B} v(r)$  where $v(r)$ is the quark-quark interaction, while the denominator is proportional to $n_B$. The resulting $ m_{\mathrm{int}}$ scales with $v(r)$. Assuming that $v(r) \propto r^a$ and $r \propto n_B^{-1/3}$ one obtains:
\begin{equation}
m_{\mathrm{int}}  \propto n_B^{-a/3}.
\end{equation}
Therefore, as expected, at low densities $m_{\mathrm{int}}$ diverges (confinement), and at asymptotically large densities it vanishes (asymptotic freedom):
\begin{align}
\lim _{n_{B} \rightarrow 0}  m_{\mathrm{int}}=\infty, \qquad 
\lim _{n_{B} \rightarrow \infty} m_{\mathrm{int}}=0.
\end{align}
Most models assume a linear quark-quark interaction which leads to $a=1$.
To keep the analysis as general as possible we assume here a simple functional form that fulfills the above requirements  \cite{Fowler:1981rp,Chakrabarty:1989bq, Chakrabarty:1991ui,Chakrabarty:1993db,Benvenuto:1989kc,Lugones:1995vg, Benvenuto:1998tx,Lugones:2002vd, Peng:1999gh,Peng:2000ff,Wang:2000dc,Yin:2008me,Xia:2014zaa,Wen:2010zz}: 
\begin{equation}
M_{i}= m_{i} + \frac{C}{n_{B}^{a/3}} ,  
\label{eq:effective_mass}
\end{equation}
where $C$ and $a$ are positive flavor independent free parameters to be determined. 

On the other hand, an effective quark mass can be derived from the pole of the resummed one-loop quark propagator at finite chemical potential which is calculated in the hard dense loop approximation (see \cite{Pisarski:1989wb, Blaizot:1993bb, Schertler:1996tq, Schertler:1997za} and references therein):
\begin{equation}
M_i=\frac{m_{i}}{2}+\sqrt{\frac{m_{i}^{2}}{4}+\frac{4 \pi \alpha_{s} \mu_{i}^{2}}{6 \pi^{2}}},
\label{eq:mass_HDL}
\end{equation}
where $m_{i}$ is the quark current mass, $\alpha_{s}$ is the strong interaction coupling constant and $\mu_{i}$ is the chemical potential of flavor $i$. 
Notice that $M_i$ grows roughly linearly with $\mu_{i}$  for $\mu_i \gg m_i$,  suggesting that the effective mass in the QMDDM  could be modified to include an additional term that increases with the density, as done in Refs. \cite{Xia:2014zaa,Chu:2021vvv,Chang:2013wnl}. The inclusion of such a term within the canonical reformulation of the QMDDM is left for a future work. However, since $M_i$ in Eq. \eqref{eq:mass_HDL} is flavor-dependent, we will analyze here the case where the quasiparticle masses are flavor dependent. To retain a qualitative behavior similar to that in Eq. \eqref{eq:effective_mass}, we will adopt a flavor dependent formula of the form: 
\begin{equation}
M_{i}= m_{i} + \frac{C}{n_{i}^{a/3}} . 
\end{equation}
%

%-------------------------------------------------------
\subsection{EOS with a flavor dependent mass formula}
\label{sec:4b}
%-------------------------------------------------------

Let us assume that the quark mass of flavor $i$ depends on the number density $n_i$ as follows: 
\begin{eqnarray}
M_i =  m_i +  \frac{C}{n_i^{a/3}} ,    \qquad  (i = u, d, s) .
\label{eq:mass_formula_uds_model1}
\end{eqnarray}
We will work in the canonical ensemble assuming that quark matter can be described as a mixture of non-interacting quarks with effective masses $M_i ~ (i=u, d, s)$ and free electrons. The Helmholtz free energy of the mixture is simply the sum of the contribution of each species:
\begin{equation}
F(V,\{n_i\}) = \sum_{i=u, d, s, e}  F_i . 
\end{equation}
For quarks,  the Helmholtz free energy is given by  Eq. \eqref{eq:Helmholtz_free_energy_2}  but with the replacement $M(n) \rightarrow M_i(n_i)$:
\begin{equation}
F_i \equiv F(V,n_i) = V g  M_i(n_i)^4 \chi(x_i) ,
\label{eq:F_flavor_dependent}
\end{equation}
where $g= 2 (\text{spin}) \times  3 (\text{color}) = 6$ and
\begin{equation}
x_i  = \frac{1}{M_i(n_i)} \left( \frac{6 \pi^2 n_i}{g} \right)^{1/3} .
\end{equation}
Electrons don't have density dependent masses (they don't perceive strong interactions); thus, their contribution is given by:
\begin{equation}
F_e = V g_e  m_e^4 \chi(x_e),
\label{eq:F_e}
\end{equation}
where $g_e=2$, $x_e = m_e^{-1} (6 \pi^2 n_e/g_e)^{1/3}$, being $m_e$ the electron's mass. 

Notice that each $F_i$ depends only on the particle number density of the same index $i$.  Therefore, all thermodynamic quantities will have the same functional form as in the one-flavor case, as we will show below.  
The energy density at $T=0$ is given by  $\epsilon = F/V$ and reads: 
\begin{equation}
\epsilon(\{n_i\}) =  \sum_{i=u, d, s, e}  {F_i}/{V}  = \sum_{i=u, d, s, e} \epsilon_i
\end{equation}
where 
\begin{equation}
\epsilon_i \equiv \epsilon_i(n_i) =
\begin{cases} g M_i^4 \chi(x_i) & (i = u, d, s) , \\ 
 g_e m_e^4 \chi(x_e)  &  (\text{electrons}) .
\end{cases}
\label{eq:62}
\end{equation}
The total pressure $p = - \partial F/\partial V|_{\{N_i \}}$ is:
\begin{eqnarray}
p &=& - \sum_i \frac{\partial F_i}{\partial V}\bigg|_{\{N_i\}} =  - \sum_i \frac{\partial (\epsilon_i V/N_i)}{\partial (V/N_i)}\bigg|_{\{N_i\}}  \\
 &=&  \sum_i n_i^2 \frac{\partial (\epsilon_i /n_i)}{\partial n_i}  = \sum_i p_i
\end{eqnarray}
being
\begin{equation}
p_i \equiv p_i(n_i) =
\begin{cases}  p_i^{FG} - B_i & (i = u, d, s) , \\ 
g_e m_e^4 \phi(x_e), &  (\text{electrons}) ,
\end{cases}
\end{equation}
where the ``free gas" contribution  $p_i^{\mathrm{FG}}$ and the ``bag constant" $B_i$ are given by:
\begin{eqnarray}
p_i^{\mathrm{FG}}(V,N_i) & \equiv &    g M_i^4 \phi(x_i), \\ 
- B_i(V,N_i) & \equiv&  g n_i  M_i^3  \frac{\partial M_i}{\partial n_i} \beta(x_i) . \label{eq:Bag_dependent}
\end{eqnarray}
Finally, the chemical potential $\mu_i$ is:
\begin{eqnarray}
\mu_i(V,N_i)  & = &   \frac{\partial F}{\partial N_i}\bigg|_{V, N_{j \neq i}} =  \frac{\partial \epsilon_i}{\partial n_i} \\
&= &  \mu_i^{\mathrm{FG}} - \frac{B_i}{n_i}  \qquad (i=u, d, s) , 
\end{eqnarray}
being 
\begin{equation}
\mu_i^{\mathrm{FG}}  = M_i \sqrt{x_i^{2}+1}.
\label{eq:mu_FG}
\end{equation}

%---------------------------------------------------------
\subsection{EOS with a flavor blind mass formula}
\label{sec:4c}
%------------------------------------------------------------

Now, let us assume that the quark mass of flavor $i$ depends on the baryon number density $n_B$ as follows: 
\begin{eqnarray}
M_i = m_i  +  \frac{C}{n_B^{a/3}} ,    \qquad  (i = u, d, s),
\label{eq:mass_formula_model2}
\end{eqnarray}
where $n_B = \tfrac{1}{3} (n_u + n_d + n_s)$.
The flavor dependence of $M_i$ comes only from the different values of the current masses $m_i$. 

Again, we will describe the system as a mixture of non-interacting quarks with effective masses $M_i$ 
and free electrons. The total Helmholtz free energy is simply the sum of the contribution of each species:
\begin{equation}
F(V,n_B) = \sum_{i=u, d, s, e}  F_i . 
\end{equation}
For quarks, $F_i$ is given by Eq. \eqref{eq:Helmholtz_free_energy_2} with the replacement $M(n) \rightarrow M_i(n_B)$:   
\begin{equation}
F_i \equiv F_i(V,\{n_i\}) = V g  M_i^4(n_B) \chi(x_i) ,
\label{eq:F_flavor_blind}
\end{equation}
where:
\begin{equation}
x_i  = \frac{1}{M_i(n_B)} \left( \frac{6 \pi^2 n_i}{g} \right)^{1/3} .
\label{eq:x_flavor_blind}
\end{equation}
Notice that in the case of quarks each $F_i$ depends on the particle number densities of all three flavors,  and the thermodynamic quantities derived from $F$ will \textit{not} necessarily have the same functional form as in the one-flavor case. 
For electrons, $F_e$ is given by Eq. \eqref{eq:F_e} and,  therefore,  $\epsilon_e$, $p_e$ and $\mu_e$ are given by the same expressions of the previous subsection.

The energy density at $T=0$ is given by  $\epsilon = F/V$ and reads: 
\begin{equation}
\epsilon(\{n_i\}) =  \sum_{i=u, d, s, e}  {F_i}/{V}  = \sum_{i=u, d, s, e} \epsilon_i ,
\label{eq:epsilon_blind}
\end{equation}
where
\begin{eqnarray}
\epsilon_i =    g M_i^4(n_B) \chi(x_i)  \qquad  (i = u, d, s).  
\label{eq:epsilon_i_blind}
\end{eqnarray}

The total pressure is given by $p = - \partial F/\partial V$  keeping ${\{N_i \}}$  constant.  Using $F = \sum_i F_i$, replacing $F_i = \epsilon_i/V$ and dividing the numerator and denominator of each term by  $N_i$ we obtain:
\begin{eqnarray}
p &=& - \sum_i \frac{\partial F_i}{\partial V}\bigg|_{\{N_i\}} =  - \sum_{i=u,d,s} \frac{\partial (\epsilon_i  \frac{V}{N_i})}{\partial (\frac{V}{N_i})}\bigg|_{\{N_i\}}  + p_e  \qquad \\
 &=&  \sum_{i=u,d,s}  n_i^2 \frac{\partial (\epsilon_i /n_i)}{\partial n_i}   + p_e  .
\end{eqnarray}
Thus, the total pressure reads  
\begin{equation}
p =  \sum_{i=u,d,s}  p_i  + p_e,
\end{equation}
being
\begin{equation}
p_i =  n_i^2 \frac{\partial (\epsilon_i /n_i)}{\partial n_i}    \qquad (i = u, d, s). 
\label{eq:p_i_blind}
\end{equation}
Replacing Eq. \eqref{eq:epsilon_i_blind} into Eq. \eqref{eq:p_i_blind} and using 
\begin{eqnarray}
\frac{\partial x_i}{\partial n_i} &=&  \frac{x_i}{3 n_i}   -  \frac{x_i}{M_i} \frac{\partial M_i}{\partial n_i} \qquad (i = u, d, s),
\label{eq:dxdni_blind} 
\end{eqnarray}
we obtain
\begin{equation}
p_i =  p_i^{FG} - B_i      \quad (i = u, d, s), 
\end{equation}
where the ``free gas" contribution  $p_i^{\mathrm{FG}}$ and the ``bag constant" $B_i$ are given by:
\begin{eqnarray}
p_i^{FG} &\equiv &   g M_i^4 \phi(x_i), \\
- B_i & \equiv &  g n_i  M_i^3  \frac{\partial M_i}{\partial n_i} \beta(x_i)  . \label{eq:Bag_blind}
\end{eqnarray}
Notice that  $p_i^{\mathrm{FG}}$ and $B_i$ have the same form in the flavor blind and in the flavor dependent cases. 

The chemical potential $\mu_i$ is:
\begin{eqnarray}
\mu_i  & = &   \frac{\partial F}{\partial N_i}\bigg|_{V, N_{j \neq i}} =  \frac{\partial \sum_k \epsilon_k}{\partial n_i} \bigg|_{n_{j \neq i}} =  \sum_k \frac{\partial \epsilon_k}{\partial n_i} \bigg|_{n_{j \neq i}}   \\
 & = & g \sum_k \frac{\partial (M^4_k \chi(x_k))}{\partial n_i} \bigg|_{n_{j \neq i}} 
\end{eqnarray}
Performing the derivative of the product and using Eq. \eqref{eq:dxdni_blind} we obtain: 
\begin{equation}
\mu_i  = \mu_i^{\mathrm{FG}} - \sum_{j}\frac{B_j}{n_j}  ,
\end{equation}
where $\mu_i^{\mathrm{FG}}$ is given by Eq. \eqref{eq:mu_FG}.

%------------------------------------------------------------
\subsection{Summary and discussion of the 3-flavor EOS}
\label{sec:summary_of_EOS}
%------------------------------------------------------------

As seen in the previous subsections, all thermodynamic quantities (with the exception of the chemical potentials) have the same form in both flavor dependent and flavor blind models, but with different formulae for the quasiparticle's mass and its derivative. For convenience, we summarize below all the formulae necessary to determine the EOS. The density dependent masses and their derivatives are:
\begin{eqnarray}
M_i  = &
\begin{cases} m_i + C n_i^{-a/3}  & \quad (\text{flavor dependent}) , \\ 
m_i +  C n_B^{-a/3}   & \quad (\text{flavor blind}) ,
\end{cases}  \\
\frac{\partial M_i}{\partial n_i}   = &
\begin{cases} - \frac{a}{3} C n_i^{-a/3 - 1}   & \quad (\text{flavor dependent}) , \\ 
- \frac{1}{3} \frac{a}{3} C n_B^{-a/3 - 1}   & \quad (\text{flavor blind}) ,
\end{cases}
\end{eqnarray}
being $a$ and $C$ positive free parameters.
The Helmholtz free energy is $F = \sum_{i=u, d, s, e}  F_i $, being:
\begin{equation}
F_i  =
\begin{cases} g V M_i^4 \chi(x_i) & \quad (i = u, d, s) , \\ 
 g_e V m_e^4 \chi(x_e)  & \quad  (\text{electrons}) ,
\end{cases}
\label{eq:F_i_summary}
\end{equation}
with $g=6$ and $g_e =2$.  The function $\chi(x)$ is defined in Eq. \eqref{eq:chi} and: 
\begin{eqnarray}
x_i  = \frac{1}{M_i} \left( \frac{6 \pi^2 n_i}{g} \right)^{1/3} & \quad (i = u, d, s) , \\ 
x_e  =  \frac{1}{m_e} \left( \frac{6 \pi^2 n_e}{g_e} \right)^{1/3} & \quad  (\text{electrons}) .
\end{eqnarray}
The energy density is $\epsilon = \sum_{i=u, d, s, e}  \epsilon_i$, where:
\begin{equation}
\epsilon_i  =
\begin{cases} g M_i^4 \chi(x_i) & \quad (i = u, d, s) , \\ 
 g_e m_e^4 \chi(x_e)  & \quad (\text{electrons}) .
\end{cases}
\label{eq:epsilon_i_summary}
\end{equation}
The total  pressure is $p = \sum_{i=u, d, s, e}  p_i$, being
\begin{equation}
p_i  =
\begin{cases} g M_i^4 \phi(x_i) - B_i  & \quad (i = u, d, s) , \\ 
  g_e m_e^4 \phi(x_e)  & \quad (\text{electrons}) ,
\end{cases}
\label{eq:p_i_summary}
\end{equation}
where $\phi(x)$ is defined in Eq. \eqref{eq:phi} and  the ``bag constant'' $B_i$ is given by: 
\begin{eqnarray}
- B_i & = &  g n_i  M_i^3  \frac{\partial M_i}{\partial n_i} \beta(x_i) ,
\label{eq:bag_summary}
\end{eqnarray}
with $\beta(x)$ defined in Eq. \eqref{eq:beta}. 
Notice that, due to the density dependence of the mass, a new term appears (the ``bag constant'') which naturally introduces confinement in the model, without the necessity of introducing an ad hoc constant. This bag constant depends on the density of the system and does not break thermodynamic consistency as in previous versions of the model. Notice that $B_i$ does not appear in the expression for energy density as in some previous versions of the QMDDM \cite{Benvenuto:1989kc,Lugones:1995vg}.

Finally, the chemical potential of $u$, $d$ and $s$ quasiparticles is: 
\begin{equation}
\mu_i  =
\begin{cases} M_i \sqrt{x_i^{2}+1} - \frac{B_i}{n_i}  &  (\text{flavor dependent}),  \\ 
M_i \sqrt{x_i^{2}+1} - \sum_{j}\frac{B_j}{n_j}  &  (\text{flavor blind})  ,
\end{cases}
\label{eq:mu_summary}
\end{equation}
while for electrons it reads:
\begin{eqnarray}
\mu_e = m_e \sqrt{x_e^{2}+1} . 
\label{eq:mu_summary_2}
\end{eqnarray}
Notice that the chemical potential in Eq. \eqref{eq:mu_summary} has two terms. The first one, $\mu^{\mathrm{FG}}$, has the same form as for the free Fermi gas and takes into account that in a completely degenerate fermionic system any new particle must be added with the Fermi energy. The second one, ${B_i}/{n_i}$ or  $\sum_{j} {B_j}/{n_j}$, arises naturally in a free Fermi gas with density dependent $M_i$. It takes into account that the energy cost of adding a new particle is smaller than in the case of a medium-independent mass (see discussion in Sec. \ref{sec:mu_1fl}). Remarkably, that new term resembles repulsive vector interactions in other approaches such as the MIT and the NJL model. However, the shift in the effective chemical potential of the above mentioned models is proportional to the density, while in the QMDDM it is always a decreasing function of the density (as already shown in Figs. \ref{fig:2}$cd$).

Finally, as remarked before, thermodynamic consistency is guaranteed in the present version of the QMDDM because we worked within the canonical ensemble, which depends naturally on $n_i$. In particular, the minimum of the energy per baryon occurs at zero pressure and the Euler's relation is verified.

%---------------------------------------------------------
\subsection{Chemical equilibrium and charge neutrality}
\label{sec:4e}
%---------------------------------------------------------

In the following, we focus on cold dense quark matter under typical NS conditions. 
Therefore, we work under the assumption that quark matter is in chemical equilibrium under weak interaction processes, such as:
\begin{equation}
\begin{aligned}
u+e^{-} & \rightarrow  d+\nu_{e}, \\
u+e^{-} & \rightarrow  s+\nu_{e},  \\
d   & \rightarrow u+e^- + \bar{\nu}_e,   \\
s &  \rightarrow u+e^- + \bar{\nu}_e,  \\
u + d &  \leftrightarrow u+s .
\end{aligned}
\end{equation}
In cold matter, neutrinos leave freely the system ($\mu_{\nu_{e}}=0$) and the chemical equilibrium conditions read: 
\begin{eqnarray}
\mu_d  & =& \mu_u + \mu_e ,  \label{eq:chemical_equilibrium_1} \\ 
\mu_s & = & \mu_d ,
\label{eq:chemical_equilibrium_2}
\end{eqnarray}
where the chemical potentials are functions of the particle number densities (cf. Eqs. \eqref{eq:mu_summary} and \eqref{eq:mu_summary_2}).

We also assume local electric charge neutrality; therefore:
\begin{equation}
\tfrac{2}{3} n_{u}-\tfrac{1}{3} n_{d}-\tfrac{1}{3} n_{s}-n_{e}=0. 
\label{eq:charge_neutrality}
\end{equation}
The three conditions above, relate the particle number densities $n_u$, $n_d$, $n_s$, and $n_e$, meaning that only one of them is independent.  

Alternatively, we can use the definition of the baryon number density:
\begin{eqnarray}
n_B = \tfrac{1}{3} (n_u + n_d + n_s)  .
\label{eq:baryon_number_density}
\end{eqnarray}
Eqs. \eqref{eq:chemical_equilibrium_1}, \eqref{eq:chemical_equilibrium_2}, \eqref{eq:charge_neutrality} and \eqref{eq:baryon_number_density} relate the variables 
$n_u$, $n_d$, $n_s$, $n_e$ and $n_B$; i.e. all thermodynamic quantities can be obtained for a given value of $n_B$.

%---------------------------------------------------------
\section{Numerical results for the EOS under compact star conditions}
\label{sec:5}
%---------------------------------------------------------

%%%%%%%%%%%%%%%   FIGURE 4
\begin{figure*}[tbh]
\centering
\includegraphics[width=0.95\columnwidth,angle=0]{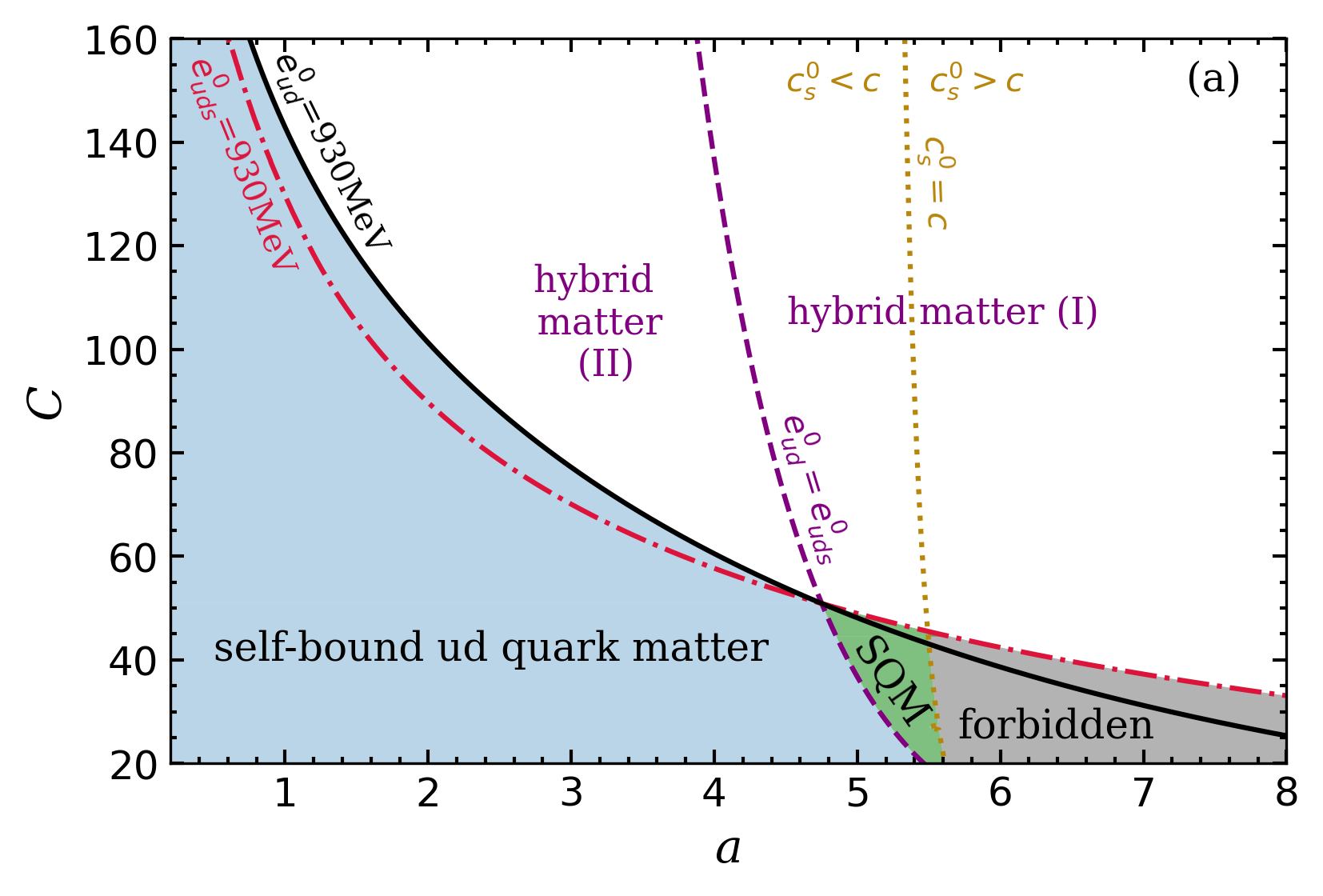}
\includegraphics[width=0.95\columnwidth,angle=0]{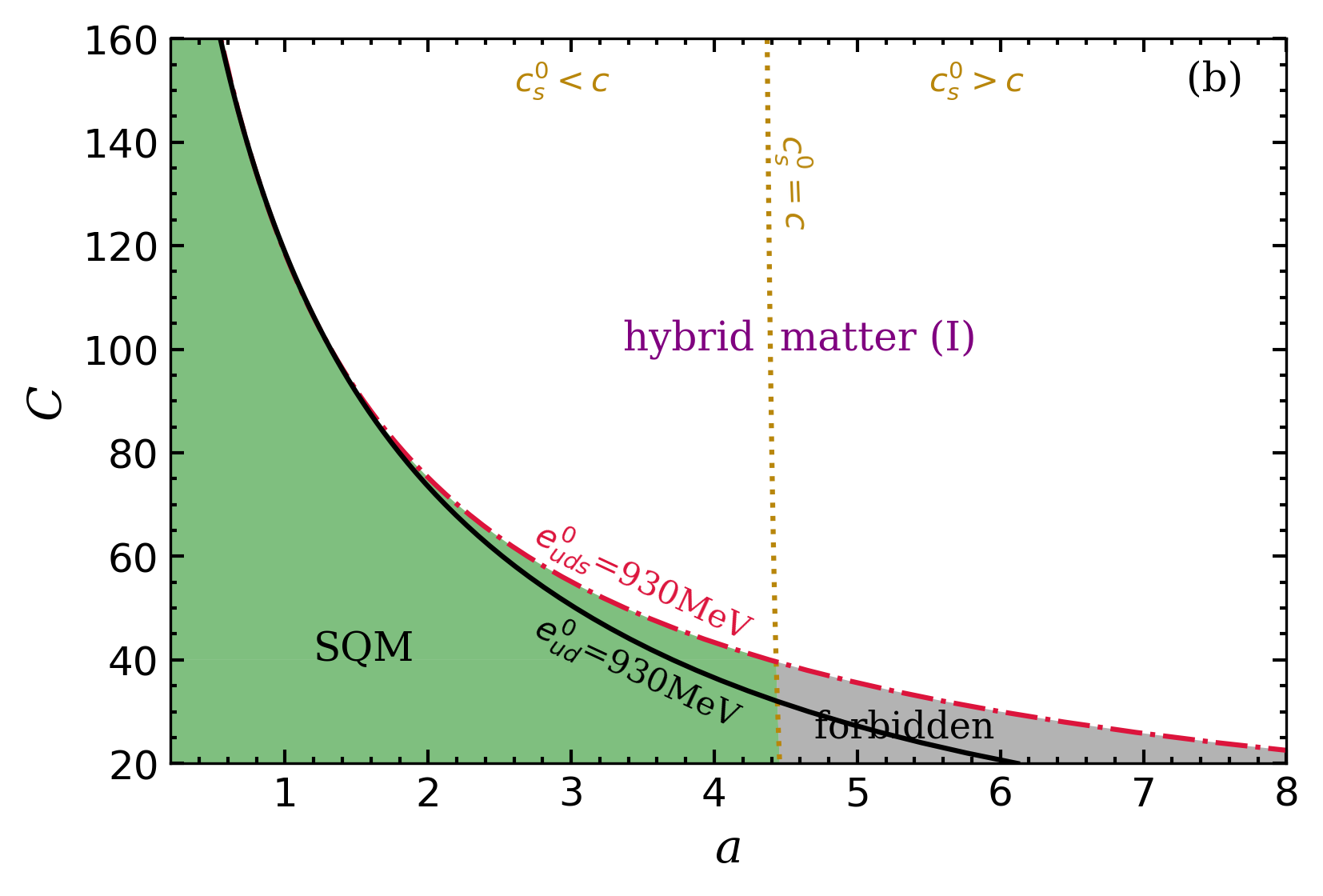}
\caption{Parameter space of quark matter in bulk for the (a) \textit{flavor dependent} and (b) \textit{flavor blind} models. If a set of parameters is chosen below the black solid line the resulting EOS will have $e^0_{ud} < 930 \mathrm{MeV}$. Similarly, if $a$ and $C$ are taken below the red dash-dotted line the EOS will have $e^0_{uds} < 930 \mathrm{MeV}$. Thus, below these lines quark matter will be sb-udQM if $e^0_{ud} < e^0_{uds}$ and SQM if  $e^0_{uds} < e^0_{ud}$. Above the $930 ~ \mathrm{MeV}$ lines, dense matter is hybrid, i.e. the preferred state is hadronic at low $p$ and deconfined at high $p$.  The label II in the hybrid matter region denotes the possibility of having a first order phase transition from $ud$ to $uds$ matter, which does not happen in case I (see Figs. \ref{fig:3fl_gibbs_dep} and   \ref{fig:3fl_EOS_dep}).   In panel (a) the curve where $e^0_{ud} = e^0_{uds}$ is shown with a dashed line; to the right of this line quark matter is 3-flavored at high enough $p$ and 2-flavored at low enough $p$. 
In panel (b) we find that the $e^0_{ud} = 930 \mathrm{MeV}$ curve is always below the $e^0_{uds} = 930 \mathrm{MeV}$ one; thus, sb-udQM and hybrid matter II are not possible. If $C$ and $a$ are chosen to the left of the $c^0_s = c$ curve, the resulting EOS will always be causal. However, to the right of that curve the EOS is acausal at low enough $p$ and self-bound matter is forbidden. }
\label{window_3flavor.png}
\end{figure*}
%%%%%%%%%%%%%%% 

%---------------------------------------------------------
\subsection{Analysis of the parameter space}
%---------------------------------------------------------

In all the following calculations we adopt fixed values for the current quark masses: $m_{u}=m_{d} = 3.4 \, \mathrm{MeV}$ and $m_{s}= 93 \, \mathrm{MeV}$ \cite{ParticleDataGroup:2020ssz}.

Let $e^0$ be the energy per baryon of \textit{bulk} quark matter at vanishing pressure and temperature: $e^0 \equiv \epsilon /n_B (p=T=0)$. Depending on the choice of the two free parameters, $a$ and $C$, $e^0$ can be greater or less than the energy per nucleon of the most tightly bound free atomic nucleus ($^{62}\mathrm{Ni}$), which is $\approx 930 ~ \mathrm{MeV}$. Therefore, we have two qualitatively different situations:
\begin{enumerate}
\item[(a)] If the parameters of the model are such that  $e^0 < 930 ~ \mathrm{MeV}$ we are in the case of self-bound quark matter. This means that \textit{bulk} quark matter in vacuum will not decay into hadronic matter, i.e. it is absolutely stable \cite{Farhi:1984qu}. If self-bound quark matter is made of only two flavors at vanishing pressure and temperature, we will designate it as \textit{self-bound $ud$ quark matter} (sb-udQM). If it is made of three flavors at $P=T=0$, we will call it \textit{strange quark matter} (SQM). If quark matter is self-bound, compact stars made of quark matter from the center up to the surface would be possible in Nature (self-bound quark stars). 

\item[(b)]  On the contrary, if $e^0  > 930 ~ \mathrm{MeV}$, the preferred state of dense matter is hadronic at low enough $p$ and deconfined at high $p$ (hybrid matter). In this case,  stars containing quark matter are always hybrid stars.  We will distinguish two types of hybrid matter: (I) when the quark phase is always made of $uds$ quarks and (II) when there is a first order phase transition between $ud$ and $uds$ matter.

\end{enumerate}

\begin{figure}[tb]
\centering
\includegraphics[width=0.97\columnwidth,angle=0]{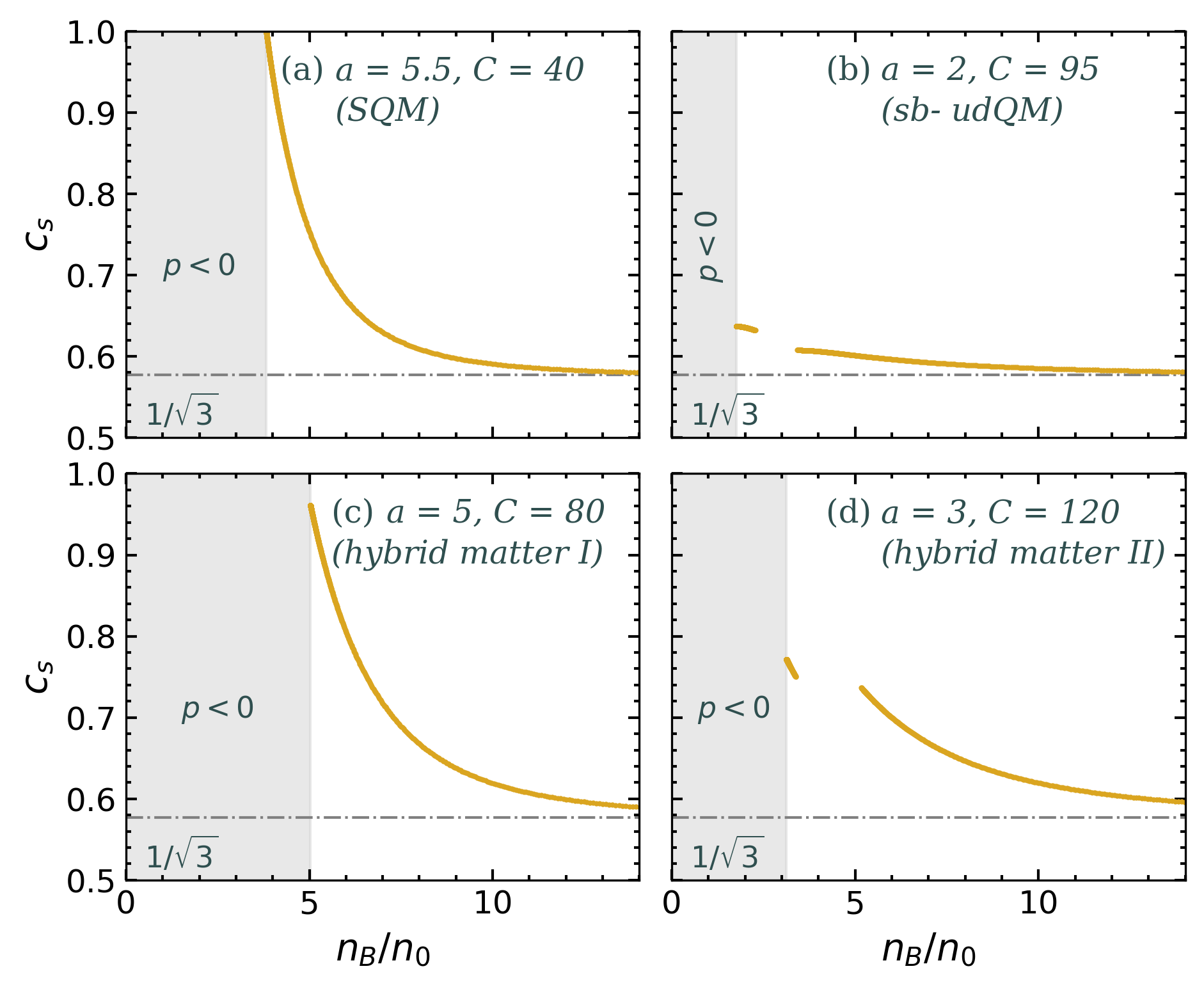}
\caption{Speed of sound in the flavor dependent model.  }
\label{fig:3fl_cs_dep}
\end{figure}

\begin{figure}[tb]
\centering
\includegraphics[width=0.99\columnwidth,angle=0]{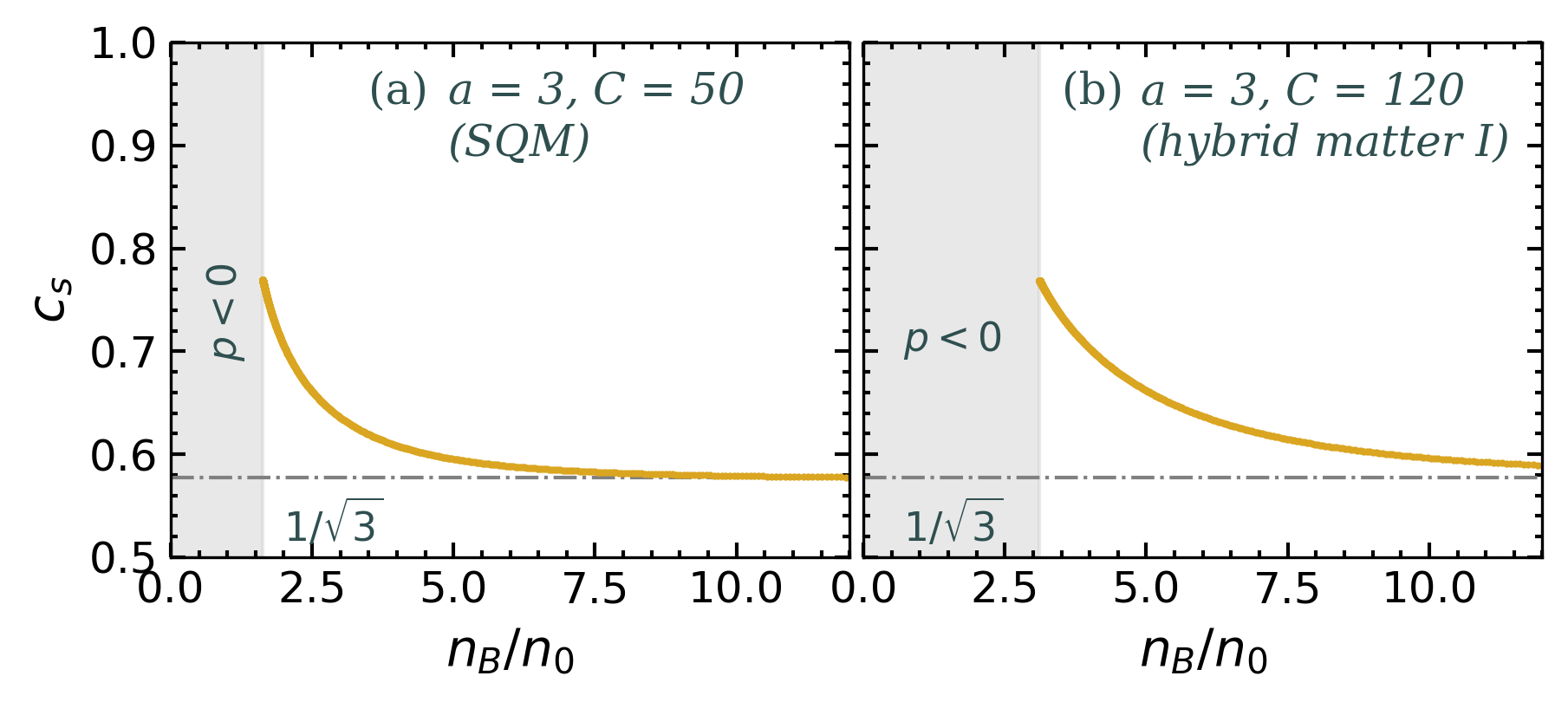}
\caption{Same as in the previous figure but for the flavor blind model.  }
\label{fig:3fl_cs_blind}
\end{figure}

In the following we analyze some general properties of the quark matter EOS that are determined by the choice of $a$ and $C$. 
To this end we will obtain the following curves in the parameter space ($C$ versus $a$ diagram):
\begin{enumerate}
\item[(1)] Curve with $e^0_{ud} = 930 ~ \mathrm{MeV}$:  This curve is obtained by imposing the following conditions:
\begin{eqnarray}
p_u + p_d + p_e = 0 , \\
(e^0_u + e^0_d + e^0_e) = 930 \, \mathrm{MeV} .
\end{eqnarray}
Together with charge neutrality and chemical equilibrium we have four equations and five unknowns ($n_u$, $n_d$, $n_e$, $C$, $a$). Thus, for each given value of $a$ we can obtain the corresponding value of $C$ that fulfills all the above conditions. This curve is shown in Fig. \ref{window_3flavor.png}  for the flavor dependent (panel $a$) and the flavor blind (panel $b$) cases presented before. Below this line $e^0_{ud} < 930 ~ \mathrm{MeV}$. This is not necessarily in contradiction with the observed existence of atomic nuclei, due to finite size effects. In fact, the decay of an atomic nucleus into a drop of deconfined quarks has an additional cost with respect to the bulk case due to the surface and curvature energy of a finite drop. For large enough surface and curvature tensions, ud-QM may be self-bound in compact stars without being in conflict with the existence of nuclei. Above this line, deconfined 2-flavor quark matter in vacuum  will decay into hadrons.

\item[(2)] Curve with $e^0_{uds} = 930 ~ \mathrm{MeV}$:  The curve is obtained by imposing:
\begin{eqnarray}
p_u + p_d + p_s + p_e = 0,  \\
(e^0_u + e^0_d +  e^0_s + e^0_e) = 930 \, \mathrm{MeV},
\end{eqnarray}
which, together with charge neutrality and chemical equilibrium, provide five equations for six unknowns ($n_u$, $n_d$, $n_s$ $n_e$, $C$, $a$). Thus, for each  $a$ one obtains the value of $C$ that fulfills all previous conditions (see Fig. \ref{window_3flavor.png}). 
Below this line  deconfined 3-flavor quark matter is self-bound and above it decays into hadrons.

\item[(3)] Curve with $e^0_{ud} = e^0_{uds}$:   This line is present only in the flavor dependent case (Fig. \ref{window_3flavor.png}$a$) and crosses the intersection between the $e^0_{ud} = 930 ~ \mathrm{MeV}$  and the $e^0_{uds} = 930 ~ \mathrm{MeV}$ curves, as expected.  
To the left of the $e^0_{ud} = e^0_{uds}$  curve, quark matter is 3-flavored at high $p$, 2-flavored at low $p$, and there is a first order phase transition between both regimes; i.e. we have hybrid matter II  (Fig. \ref{fig:3fl_gibbs_dep}$d$) and  sb-udQM (Fig. \ref{fig:3fl_gibbs_dep}$b$). To the right of this curve bulk quark matter is 3-flavored at any pressure;  i.e. we have hybrid matter I (Fig. \ref{fig:3fl_gibbs_dep}$c$) and SQM (Fig. \ref{fig:3fl_gibbs_dep}$a$).
In the flavor blind model the $e^0_{ud} = 930 \mathrm{MeV}$ curve is always below the $e^0_{uds} = 930 \mathrm{MeV}$ one, and these lines tend to overlap for small enough values of $a$. As a consequence, sb-udQM and hybrid matter II are not possible in panel (b) of Fig. \ref{window_3flavor.png}.

\item[(4)] Curve with $c_s^0 = c$: The dotted line  in Fig. \ref{window_3flavor.png} indicates the values of $C$ and $a$ for which the speed of sound at vanishing pressure and  temperature ($c_s^0 \equiv c_s(p=T=0)$) equals the speed of light $c$. Since $c_s$ is a decreasing function of $p$ (see Figs. \ref{fig:3fl_cs_dep} and \ref{fig:3fl_cs_blind}), the region to the left of this curve is always causal. To the right, the EOS is acausal at low enough $p$. This is not necessarily a problem in the case of hybrid matter, provided that the hadronic phase is energetically favorable in the region where quark matter is acausal. However, such behavior is not admissible for self-bound matter because it exists all the way down to $p=0$. Thus, the gray region to the right of the $c^0_s = c$ curve is forbidden since the resulting EOS will always have an acausal regime at low enough $p$. 
Notice that the curve $c_s = c$ is almost vertical, meaning that the causality boundary depends only on the parameter $a$, i.e. on the intensity of the quark-quark interaction (see Sec. \ref{sec:4a}).  Notice that for the typical value $a=1$ the model is comfortably within the causal region.  Problems with causality arise for very strong quark-quark interactions with $a$ beyond $\sim 4-5$.

\end{enumerate}

All the previous curves define different regions in the parameter space that are shown and discussed in Fig. \ref{window_3flavor.png}.

%---------------------------------------------------------
\subsection{Equation of state}
%---------------------------------------------------------

%%%%%%%%%%%%%%%   FIGURE 5 + 6
\begin{figure}[tb]
\centering
\includegraphics[width=0.99\columnwidth,angle=0]{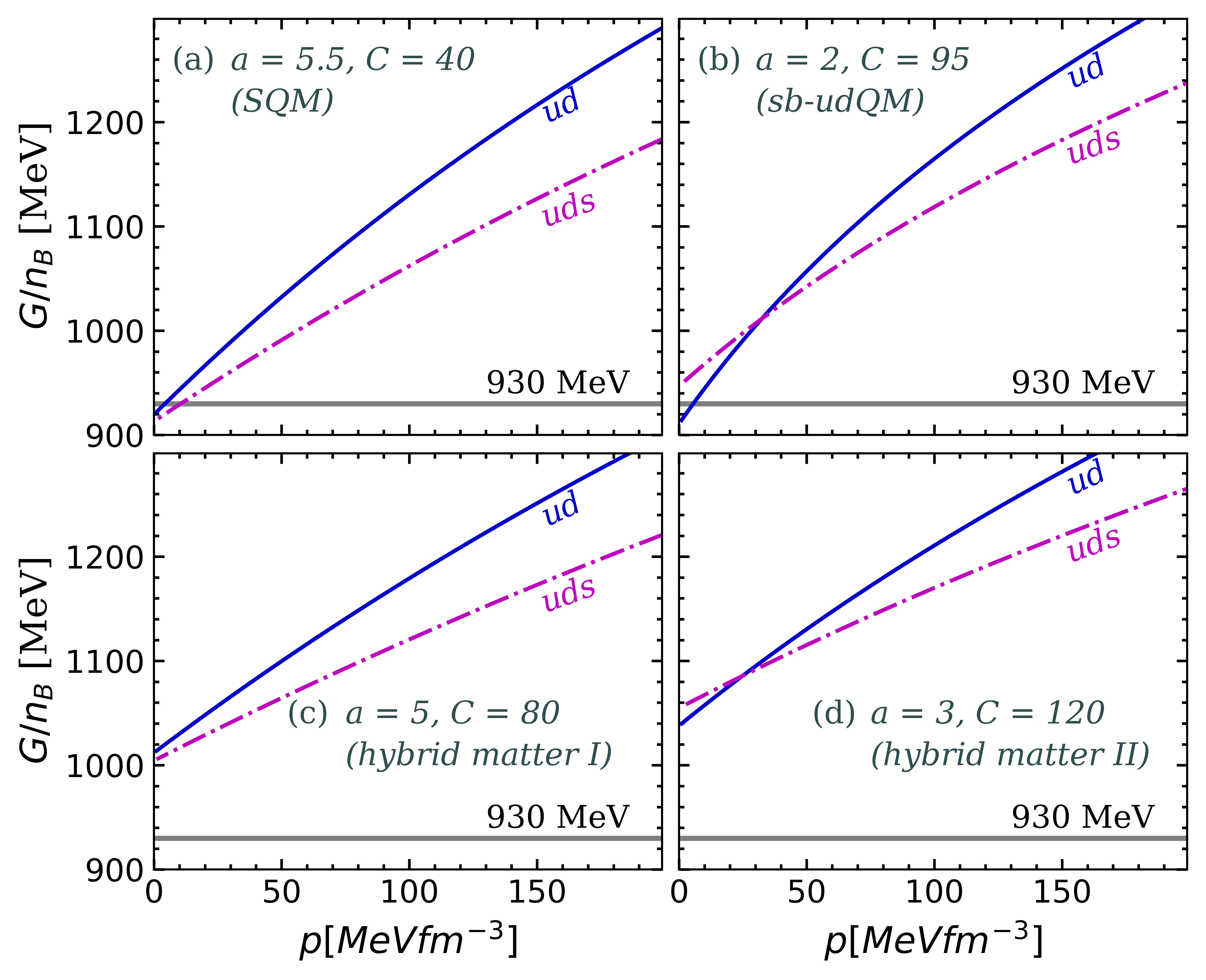}
\caption{Gibbs free energy per baryon of 2-flavor and 3-flavor quark matter in bulk for the flavor dependent case. Labels (a-d) correspond to different parameter choices that are specified within each panel. }
\label{fig:3fl_gibbs_dep}
\end{figure}

\begin{figure}[tb]
\centering
\includegraphics[width=0.99\columnwidth,angle=0]{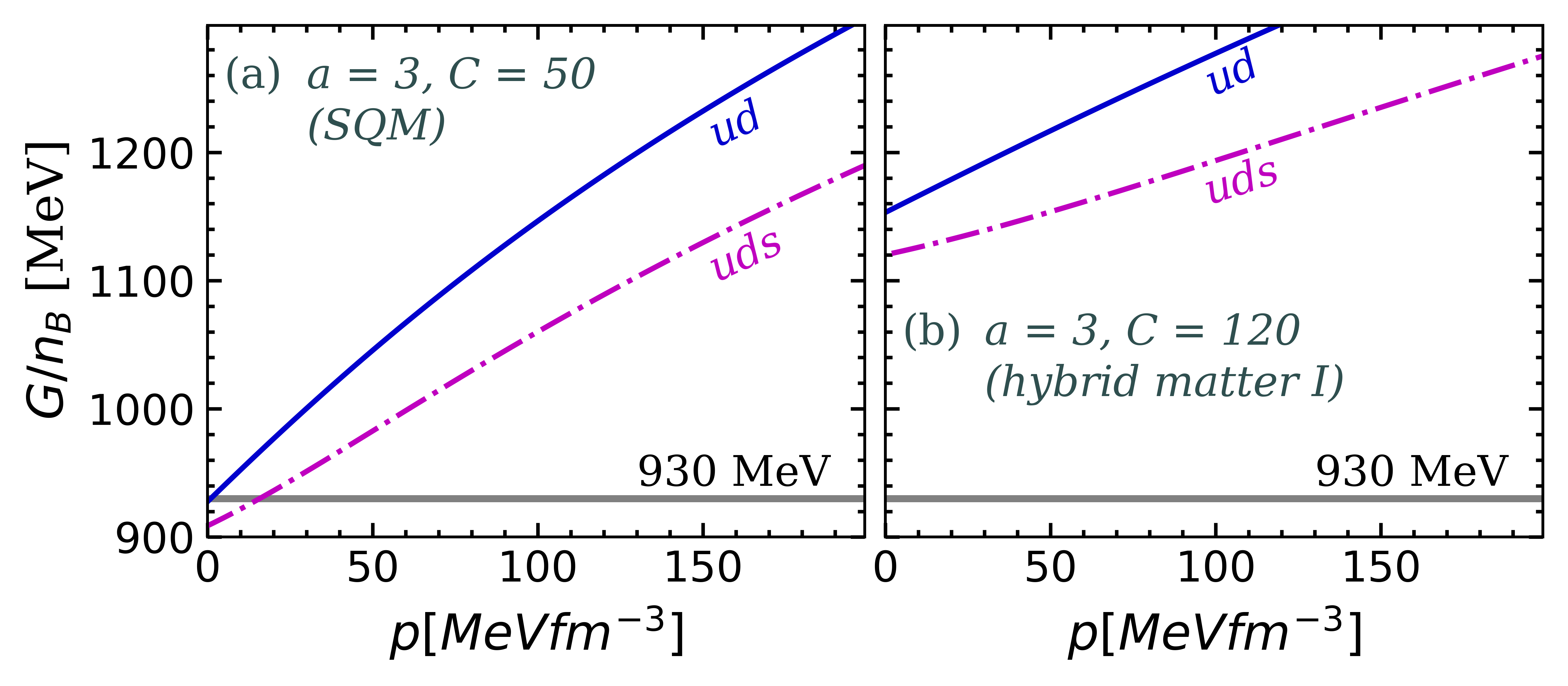}
\caption{Same as in the previous figure but for the flavor blind model.  }
\label{fig:3fl_gibbs_blind}
\end{figure}
%%%%%%%%%%%%%%%  

%%%%%%%%%%%%%%%   FIGURE 7 + 8
\begin{figure}[tb]
\centering
\includegraphics[width=0.97\columnwidth,angle=0]{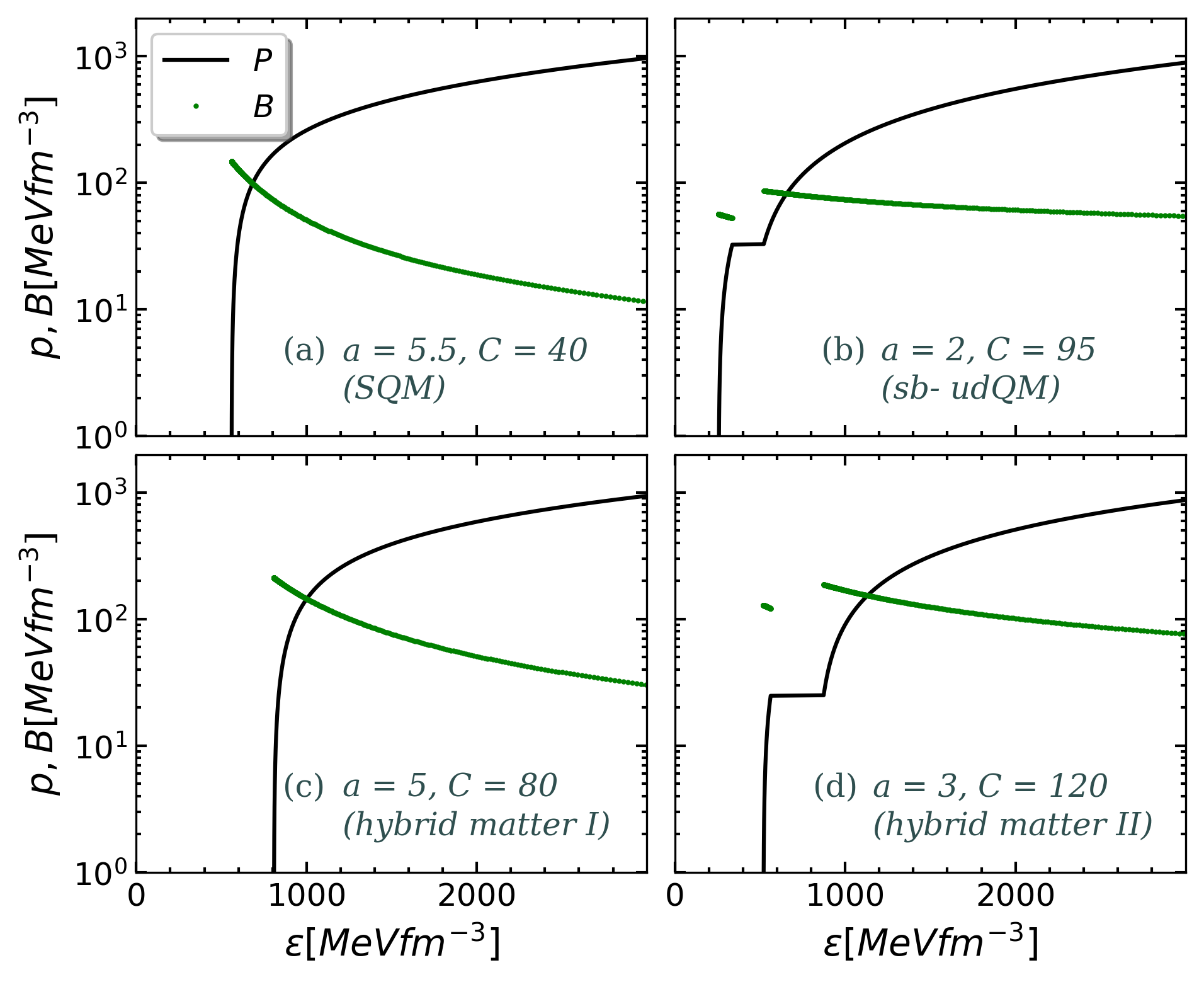}
\caption{Pressure and bag constant in the flavor dependent model. The  discontinuity of the bag constant in panels (b) and (d) occurs because the system does not exist in the density range corresponding to the plateau in the pressure. } 
\label{fig:3fl_EOS_dep}
\end{figure}

\begin{figure}[tb]
\centering
\includegraphics[width=0.99\columnwidth,angle=0]{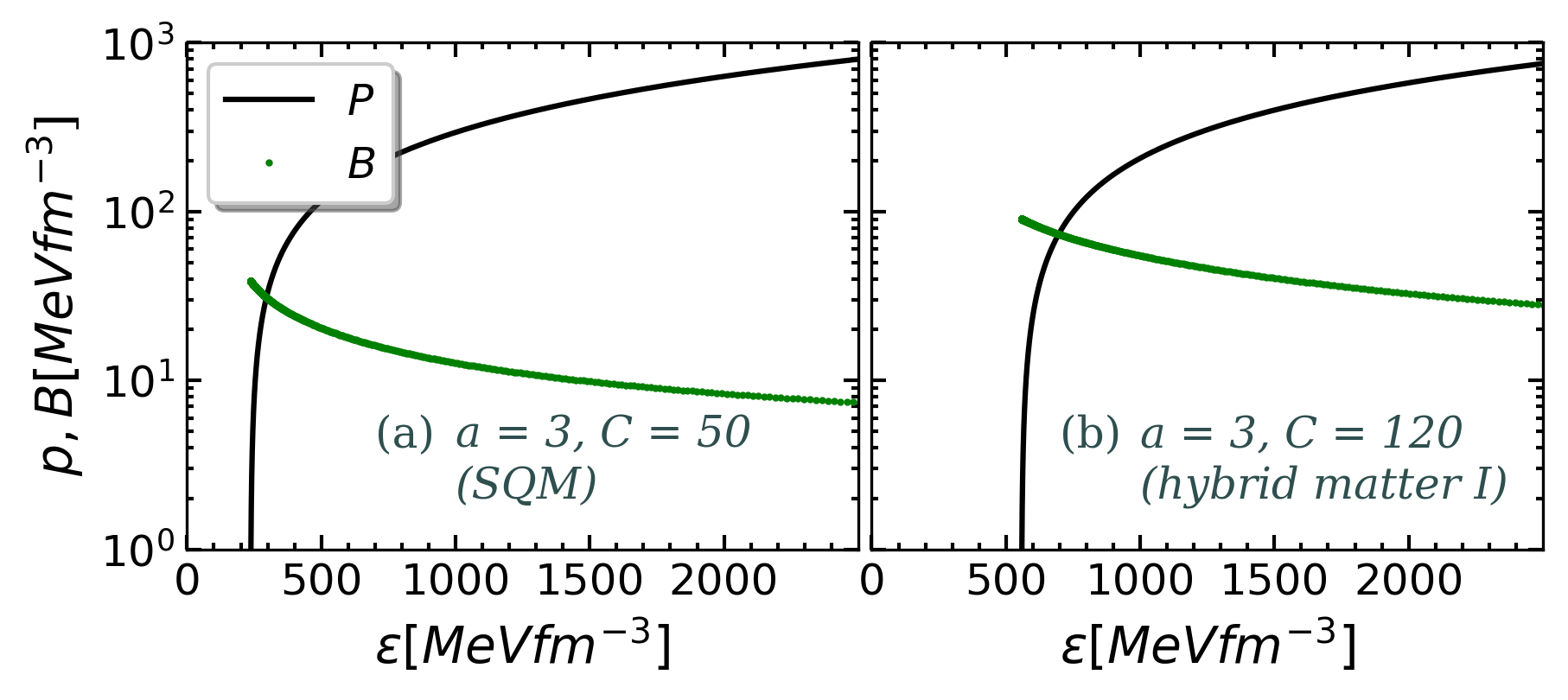}
\caption{Same as in the previous figure but for the flavor blind model.  }
\label{fig:3fl_EOS_blind}
\end{figure}
%%%%%%%%%%%%%%% 

To better understand the parameter space, we show in Figs. \ref{fig:3fl_gibbs_dep} and \ref{fig:3fl_gibbs_blind} the Gibbs free energy per baryon $G/n_B \equiv (\epsilon + p) /n_B$ as a function of the pressure. We consider parameters within all regions of Fig. \ref{window_3flavor.png}. 

First notice that, since $e^0 = G/n_B(P=T=0)$, a direct connection can be established between Fig. \ref{window_3flavor.png} and the zero pressure points of the curves in Figs. \ref{fig:3fl_gibbs_dep} and \ref{fig:3fl_gibbs_blind}. Indeed,  for SQM the magenta dash-dotted curve is always below $930 ~  \mathrm{MeV}$ at $P=0$ (Figs. \ref{fig:3fl_gibbs_dep}$a$ and \ref{fig:3fl_gibbs_blind}$a$). Also, for sb-udQM the continuous blue curve is below the $930 ~  \mathrm{MeV}$ line at $P=0$ (Fig. \ref{fig:3fl_gibbs_dep}$b$). On the contrary, for hybrid matter the value of  $G/n_B$ at $P=0$ is always above $930 ~  \mathrm{MeV}$ (Figs. \ref{fig:3fl_gibbs_dep}$cd$ and \ref{fig:3fl_gibbs_blind}$b$).

Now, let us analyze  the flavor composition of quark matter at finite pressures. In the flavor blind model, the $uds$ curve is always below the $ud$ one, as exemplified in Fig. \ref{fig:3fl_gibbs_blind}. Since the preferred phase is the one with lower $G/n_B$, one concludes that sb-udQM and hybrid matter II are not possible in this case.  
Now let us focus on the flavor dependent model. For parameter choices within the hybrid matter II and the sb-udQM regions of Fig. \ref{window_3flavor.png}a,  one finds crosses between the $ud$ and $uds$ curves
indicating that there is a first order phase transition from 2 to 3 flavors at that point (Figs. \ref{fig:3fl_gibbs_dep}$bd$). 
For choices within the hybrid matter I and the SQM regions of Fig. \ref{window_3flavor.png}a, the $uds$ curve is always below the $ud$ one, meaning that quark matter is always made of three flavors, whether self-bound or hybrid  (Figs. \ref{fig:3fl_gibbs_dep}$ac$).

In Figs. \ref{fig:3fl_EOS_dep} and  \ref{fig:3fl_EOS_blind} we show the total pressure $p= \sum_i p_i$ (Eq. \eqref{eq:p_i_summary}) and the bag constant $B = \sum_i B_i$ as a function of the energy density for the same parametrizations presented in Figs. \ref{fig:3fl_gibbs_dep} and \ref{fig:3fl_gibbs_blind}. In all cases the pressure becomes negative at finite energy density due to the effect of $B$. The bag constant depends on density, always being a decreasing function of $\epsilon$. At asymptotically large densities $B$ tends to zero and the system behaves as a free Fermi gas of electrons and quarks with $M_i = m_i$.
In Figs. \ref{fig:3fl_EOS_dep}$bd$ the pressure presents a plateau corresponding to the phase transition from 2 to 3 flavors mentioned above. For that reason, the bag constant is discontinuous in these cases.

%%%%%%%%%%%%%%%   FIGURE 9 + 10

%%%%%%%%%%%%%%% 

%---------------------------------------------------------
\subsection{Speed of sound }
%---------------------------------------------------------

The speed of sound $c_s$ is shown in Figs. \ref{fig:3fl_cs_dep} and \ref{fig:3fl_cs_blind} for the same parameter choices of previous figures.
In all cases $c_s$ is a decreasing function of the baryon number density and tends asymptotically to the conformal limit  $c_s = 1/\sqrt{3}$.
Since the pressure tends to zero at a finite density, we show a shaded region corresponding to $p<0$. For some hybrid matter parametrizations (not shown in  Figs. \ref{fig:3fl_cs_dep} and \ref{fig:3fl_cs_blind}) we find that $c_s > c$ outside the shaded region. As mentioned before, this is not necessarily a problem in the case of hybrid matter, provided that the hadronic phase is energetically favorable in the region where quark matter is acausal. 
Notice that $c_s$ is not defined in a range of densities of Figs. \ref{fig:3fl_cs_dep}$bd$ because there is a first-order phase transition between $ud$ and $uds$ matter in this region.

Finally, we find that $c_s$ grows at low densities as in Refs. \cite{Benvenuto:1989kc,Lugones:1995vg}, unlike  other versions of the QMDDM  where $c_s$ decreases at low densities \cite{Peng:1999gh,Xia:2014zaa}.  
This behavior is related to the fact that quasiparticles acquire a significant mass at low densities leading to a rapid pressure loss. This rapid variation produces a strong growth of the speed of sound $c_s^2 = {\partial p}/{\partial \epsilon}$. For extremely strong quark-quark interactions $c_s$ becomes acausal at low densities, but this does not occur for typical values of $a$. This behavior can be seen in Figs. \ref{fig:3fl_cs_dep}$a$ where, for $a=5.5$, $c_s \rightarrow c$ as $p \rightarrow 0$.

%---------------------------------------------------------
\section{Summary and Conclusions}
\label{sec:6}
%---------------------------------------------------------

In this paper we revisited the QMDDM and showed that thermodynamic inconsistencies that have plagued the model for decades, can be solved if the model is formulated in the canonical instead of the grand canonical ensemble. 

We first focused on a simple one-component system assuming that the phenomenological mass is given by
\begin{eqnarray}
M(n) = m + \frac{C}{n^{a/3}} 
\end{eqnarray}
where $n$ is the particle number density, $m$  the current mass, and $a$ and $C$ are positive free parameters. 
This functional form has been widely used in previous versions of the model because $M$ diverges for densities approaching zero  thus mimicking confinement and tends to the current quark mass resembling the restoration of the chiral symmetry in a phenomenological way when $m=0$. 
Differently from some previous works that consider $a=1$ or $a=3$, we didn't impose a priori restrictions on $a$. 
With this ansatz for the mass we  showed that a  ``bag'' term that produces quark confinement naturally appears in the pressure (and not in the energy density) due to density dependence of the quark masses. Additionally, the chemical potential gains a new term that resembles quark repulsive interactions.
Unlike some previous versions of the model, within the new formulation,  the minimum of the energy per baryon occurs at zero pressure, and  Euler's relationship is verified.

Then, we extended the formalism to the astrophysically realistic case of charge-neutral three-flavor quark matter in equilibrium under weak interactions, focusing on two different mass formulae: a flavor dependent and a flavor blind one:
\begin{eqnarray}
M_i  = &
\begin{cases} m_i +  C n_i^{-a/3}  & \quad (\text{flavor dependent}) , \\ 
m_i +   C n_B^{-a/3}  & \quad (\text{flavor blind}) .
\end{cases} 
\end{eqnarray}
For these two models, we derived the equation of state, which is summarized in Sec. \ref{sec:summary_of_EOS}.
We systematically analyzed the parameter space and identified different regions corresponding to self-bound 2-flavor and 3-flavor quark matter, hybrid matter and causal behavior (see Fig. \ref{window_3flavor.png}).
In the flavor blind model we found that there is no first order phase transition between $ud$ and $uds$ quark matter. This means that in this case we only find strange quark matter or the so called hybrid matter I. However, in a more realistic model where the $C$ parameter for the $s$-quark mass is different from the one for $u$ and $d$ flavors, the $s$ quarks disappear smoothly at both finite baryon number density and pressure (see Appendix \ref{appendix}). 
In the flavor dependent model the parameter space is richer. In fact,  a first order phase transition between $ud$ and $uds$ quark matter is possible for some parameter choices. 
We found a large region of the parameter space that allows for the existence of self-bound matter, where a small part of it corresponds to strange quark matter and the rest to self-bound $ud$ quark matter. The remainder of the parameter space corresponds to the so called hybrid matter I (with $ud-uds$ first order transition) or hybrid matter II (without a $ud-uds$ plateau).
If one focuses on the cases $a=1$ or $a=3$ adopted in many previous works, then deconfined quark matter is either hybrid matter II or self-bound $ud$ for the flavor dependent model, and  hybrid matter I or strange quark matter for the flavor blind case.
We also studied the speed of sound showing that it tends to the conformal limit  $c_s = 1/\sqrt{3}$ at asymptotically large densities. Al low densities $c_s$ is always larger that $1/\sqrt{3}$ and it may exceed the speed of light for some parameter choices. In Fig. \ref{window_3flavor.png} we show the regions where causality is always verified.

The main purpose of the present work has been to carry out a systematic and exhaustive analysis of the QMDDM in a proper statistical ensemble that allows solving long persistent thermodynamic inconsistencies of the model.
The reformulated version of the QMDDM is quite interesting because it encodes in a phenomenological way key properties of quark matter such as confinement, asymptotic freedom and chiral symmetry breaking/restoration in a simple analytical EOS that may be useful for astrophysical applications.

\subsection*{Acknowledgements}
G.L. acknowledges the support of the Brazilian agencies Conselho Nacional de Desenvolvimento Cient\'{\i}fico e Tecnol\'ogico (grant 316844/2021-7) and Funda{\c c}\~ao de Amparo \`a
Pesquisa do Estado de S\~ao Paulo  (grants 2022/02341-9 and 2013/10559-5).
A. G. G. would like to acknowledge the financial support from CONICET under Grant No. PIP 22-24 11220210100150CO,  ANPCyT (Argentina) under Grant PICT20-01847, and the National University of La Plata (Argentina), Project No. X824.

\appendix

\section{On the two to three flavor transition}
\label{appendix}

%%%%%%%%%%%%%%%   FIGURE 
\begin{figure}[t]
\centering
\includegraphics[width=0.97\columnwidth,angle=0]{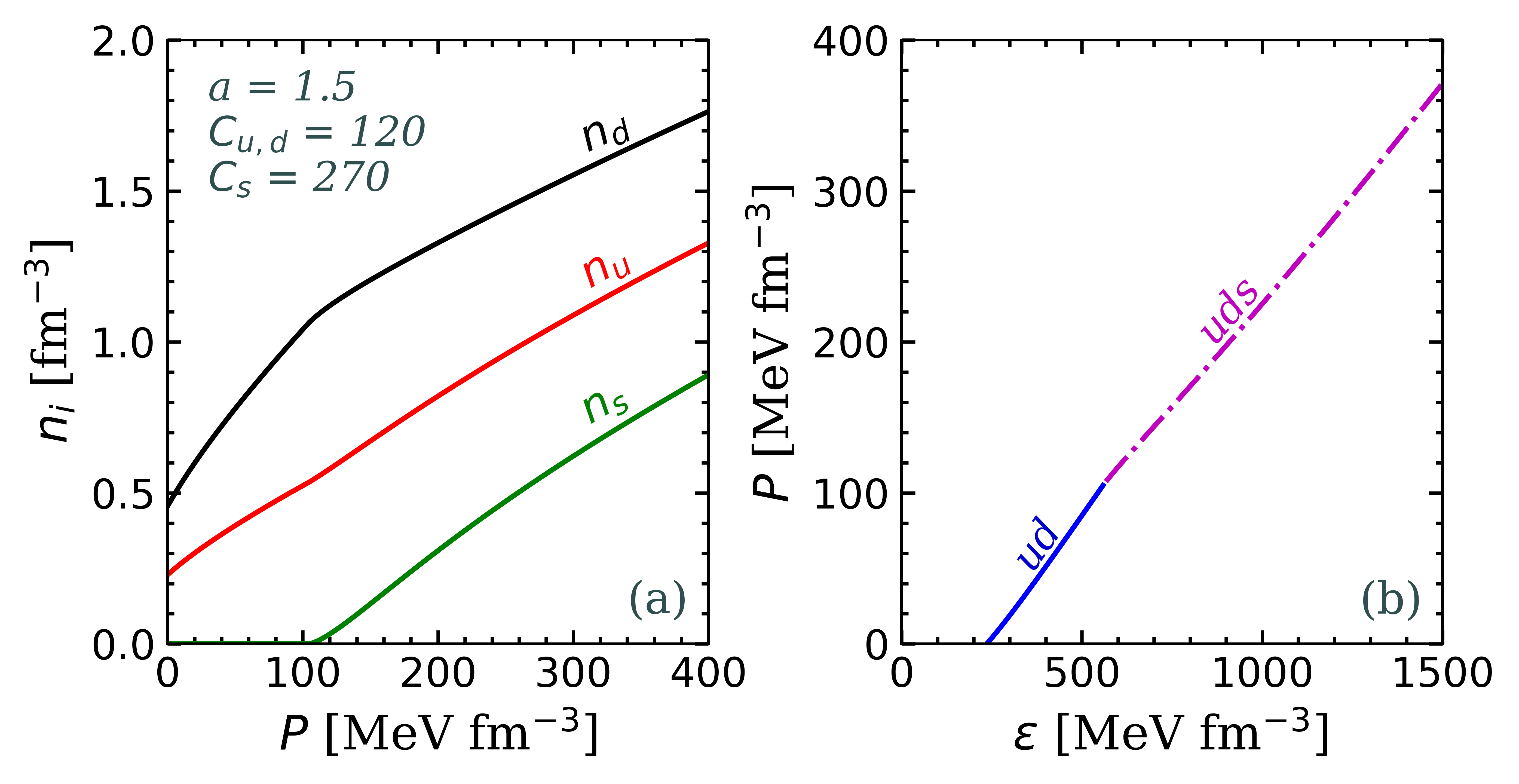}
\caption{A flavor blind model with $C_s > C_{u,d}$ results in a smooth transition from two to thee flavor quark matter. Notice that $s$ quarks disappear from the system at the same pressure at which the EOS of 2 and 3 flavors merge (cf. panels (a) and (b)).}
\label{fig:appendix}
\end{figure}

In the flavor dependent model we find that the number density of $s$ quarks vanishes at negative pressures or at a positive $p$ that is always below the pressure of the crossing between $(G/n_B)_{ud}$ and  $(G/n_B)_{uds}$. For example, for the parametrization shown in Fig. \ref{fig:3fl_gibbs_dep}b the $ud$ and $uds$ curves cross at a pressure  $p_{tr} \sim 35 ~ \mathrm{MeV fm^{-3}}$ and the number density of $s$-quarks goes to zero at a $p<0$. Since below $p_{tr}$ we find $(G/n_B)_{ud} < (G/n_B)_{uds}$, we conclude that the system would prefer to change abruptly to 2 flavors at low pressures. This situation persists even in a more realistic model where the $C$ parameter of the $s$-quark mass is different from the one for $u$ and $d$ flavors.

In the flavor blind model, when the $C$ parameter is the same for all 3 flavors we find that the $s$ quark density always vanishes at a negative pressure. However, in contrast with the flavor dependent model, when a larger $C$ is adopted for strange quarks, $n_s$ goes to zero at finite pressures and the transition from two to three flavors is smooth. This can be sen in Fig. \ref{fig:appendix} where we considered the parametrization $a=1.5$, $C_{u, d} = 120$ and $C_s = 270$. Notice that $s$ quarks disappear at a $p = 105 ~ \mathrm{MeV ~fm^{-3}}$ (panel a), where the pressure and energy density of the two and three flavor solutions coincide (panel b).

\bibliography{references}

\begin{thebibliography}{28}%
\makeatletter
\providecommand \@ifxundefined [1]{%
 \@ifx{#1\undefined}
}%
\providecommand \@ifnum [1]{%
 \ifnum #1\expandafter \@firstoftwo
 \else \expandafter \@secondoftwo
 \fi
}%
\providecommand \@ifx [1]{%
 \ifx #1\expandafter \@firstoftwo
 \else \expandafter \@secondoftwo
 \fi
}%
\providecommand \natexlab [1]{#1}%
\providecommand \enquote  [1]{``#1''}%
\providecommand \bibnamefont  [1]{#1}%
\providecommand \bibfnamefont [1]{#1}%
\providecommand \citenamefont [1]{#1}%
\providecommand \href@noop [0]{\@secondoftwo}%
\providecommand \href [0]{\begingroup \@sanitize@url \@href}%
\providecommand \@href[1]{\@@startlink{#1}\@@href}%
\providecommand \@@href[1]{\endgroup#1\@@endlink}%
\providecommand \@sanitize@url [0]{\catcode `\\12\catcode `\$12\catcode
  `\&12\catcode `\#12\catcode `\^12\catcode `\_12\catcode `\%12\relax}%
\providecommand \@@startlink[1]{}%
\providecommand \@@endlink[0]{}%
\providecommand \url  [0]{\begingroup\@sanitize@url \@url }%
\providecommand \@url [1]{\endgroup\@href {#1}{\urlprefix }}%
\providecommand \urlprefix  [0]{URL }%
\providecommand \Eprint [0]{\href }%
\providecommand \doibase [0]{https://doi.org/}%
\providecommand \selectlanguage [0]{\@gobble}%
\providecommand \bibinfo  [0]{\@secondoftwo}%
\providecommand \bibfield  [0]{\@secondoftwo}%
\providecommand \translation [1]{[#1]}%
\providecommand \BibitemOpen [0]{}%
\providecommand \bibitemStop [0]{}%
\providecommand \bibitemNoStop [0]{.\EOS\space}%
\providecommand \EOS [0]{\spacefactor3000\relax}%
\providecommand \BibitemShut  [1]{\csname bibitem#1\endcsname}%
\let\auto@bib@innerbib\@empty
%</preamble>
\bibitem [{\citenamefont {Annala}\ \emph {et~al.}(2020)\citenamefont {Annala},
  \citenamefont {Gorda}, \citenamefont {Kurkela}, \citenamefont {N\"attil\"a},\
  and\ \citenamefont {Vuorinen}}]{Annala:2019puf}%
  \BibitemOpen
  \bibfield  {author} {\bibinfo {author} {\bibfnamefont {E.}~\bibnamefont
  {Annala}}, \bibinfo {author} {\bibfnamefont {T.}~\bibnamefont {Gorda}},
  \bibinfo {author} {\bibfnamefont {A.}~\bibnamefont {Kurkela}}, \bibinfo
  {author} {\bibfnamefont {J.}~\bibnamefont {N\"attil\"a}},\ and\ \bibinfo
  {author} {\bibfnamefont {A.}~\bibnamefont {Vuorinen}},\ }\bibfield  {title}
  {\bibinfo {title} {{Evidence for quark-matter cores in massive neutron
  stars}},\ }\href {https://doi.org/10.1038/s41567-020-0914-9} {\bibfield
  {journal} {\bibinfo  {journal} {Nature Phys.}\ }\textbf {\bibinfo {volume}
  {16}},\ \bibinfo {pages} {907} (\bibinfo {year} {2020})},\ \Eprint
  {https://arxiv.org/abs/1903.09121} {arXiv:1903.09121 [astro-ph.HE]}
  \BibitemShut {NoStop}%
\bibitem [{\citenamefont {Fowler}\ \emph {et~al.}(1981)\citenamefont {Fowler},
  \citenamefont {Raha},\ and\ \citenamefont {Weiner}}]{Fowler:1981rp}%
  \BibitemOpen
  \bibfield  {author} {\bibinfo {author} {\bibfnamefont {G.~N.}\ \bibnamefont
  {Fowler}}, \bibinfo {author} {\bibfnamefont {S.}~\bibnamefont {Raha}},\ and\
  \bibinfo {author} {\bibfnamefont {R.~M.}\ \bibnamefont {Weiner}},\ }\bibfield
   {title} {\bibinfo {title} {{Confinement and Phase Transitions}},\ }\href
  {https://doi.org/10.1007/BF01410668} {\bibfield  {journal} {\bibinfo
  {journal} {Z. Phys. C}\ }\textbf {\bibinfo {volume} {9}},\ \bibinfo {pages}
  {271} (\bibinfo {year} {1981})}\BibitemShut {NoStop}%
\bibitem [{\citenamefont {Chakrabarty}\ \emph {et~al.}(1989)\citenamefont
  {Chakrabarty}, \citenamefont {Raha},\ and\ \citenamefont
  {Sinha}}]{Chakrabarty:1989bq}%
  \BibitemOpen
  \bibfield  {author} {\bibinfo {author} {\bibfnamefont {S.}~\bibnamefont
  {Chakrabarty}}, \bibinfo {author} {\bibfnamefont {S.}~\bibnamefont {Raha}},\
  and\ \bibinfo {author} {\bibfnamefont {B.}~\bibnamefont {Sinha}},\ }\bibfield
   {title} {\bibinfo {title} {{Strange Quark Matter and the Mechanism of
  Confinement}},\ }\href {https://doi.org/10.1016/0370-2693(89)90166-4}
  {\bibfield  {journal} {\bibinfo  {journal} {Phys. Lett. B}\ }\textbf
  {\bibinfo {volume} {229}},\ \bibinfo {pages} {112} (\bibinfo {year}
  {1989})}\BibitemShut {NoStop}%
\bibitem [{\citenamefont {Chakrabarty}(1991)}]{Chakrabarty:1991ui}%
  \BibitemOpen
  \bibfield  {author} {\bibinfo {author} {\bibfnamefont {S.}~\bibnamefont
  {Chakrabarty}},\ }\bibfield  {title} {\bibinfo {title} {{Equation of state of
  strange quark matter and strange star}},\ }\href
  {https://doi.org/10.1103/PhysRevD.43.627} {\bibfield  {journal} {\bibinfo
  {journal} {Phys. Rev. D}\ }\textbf {\bibinfo {volume} {43}},\ \bibinfo
  {pages} {627} (\bibinfo {year} {1991})}\BibitemShut {NoStop}%
\bibitem [{\citenamefont {Chakrabarty}(1993)}]{Chakrabarty:1993db}%
  \BibitemOpen
  \bibfield  {author} {\bibinfo {author} {\bibfnamefont {S.}~\bibnamefont
  {Chakrabarty}},\ }\bibfield  {title} {\bibinfo {title} {{Stability of strange
  quark matter at T not = 0}},\ }\href
  {https://doi.org/10.1103/PhysRevD.48.1409} {\bibfield  {journal} {\bibinfo
  {journal} {Phys. Rev. D}\ }\textbf {\bibinfo {volume} {48}},\ \bibinfo
  {pages} {1409} (\bibinfo {year} {1993})}\BibitemShut {NoStop}%
\bibitem [{\citenamefont {Benvenuto}\ and\ \citenamefont
  {Lugones}(1995)}]{Benvenuto:1989kc}%
  \BibitemOpen
  \bibfield  {author} {\bibinfo {author} {\bibfnamefont {O.~G.}\ \bibnamefont
  {Benvenuto}}\ and\ \bibinfo {author} {\bibfnamefont {G.}~\bibnamefont
  {Lugones}},\ }\bibfield  {title} {\bibinfo {title} {{Strange matter equation
  of state in the quark mass density dependent model}},\ }\href
  {https://doi.org/10.1103/PhysRevD.51.1989} {\bibfield  {journal} {\bibinfo
  {journal} {Phys. Rev.}\ }\textbf {\bibinfo {volume} {D51}},\ \bibinfo {pages}
  {1989} (\bibinfo {year} {1995})}\BibitemShut {NoStop}%
%%CITATION = PHRVA,D51,1989;%%
\bibitem [{\citenamefont {Lugones}\ and\ \citenamefont
  {Benvenuto}(1995)}]{Lugones:1995vg}%
  \BibitemOpen
  \bibfield  {author} {\bibinfo {author} {\bibfnamefont {G.}~\bibnamefont
  {Lugones}}\ and\ \bibinfo {author} {\bibfnamefont {O.~G.}\ \bibnamefont
  {Benvenuto}},\ }\bibfield  {title} {\bibinfo {title} {{Strange matter
  equation of state and the combustion of nuclear matter into strange matter in
  the quark mass density dependent model at $T > 0$}},\ }\href
  {https://doi.org/10.1103/PhysRevD.52.1276} {\bibfield  {journal} {\bibinfo
  {journal} {Phys. Rev.}\ }\textbf {\bibinfo {volume} {D52}},\ \bibinfo {pages}
  {1276} (\bibinfo {year} {1995})}\BibitemShut {NoStop}%
%%CITATION = PHRVA,D52,1276;%%
\bibitem [{\citenamefont {Benvenuto}\ and\ \citenamefont
  {Lugones}(1998)}]{Benvenuto:1998tx}%
  \BibitemOpen
  \bibfield  {author} {\bibinfo {author} {\bibfnamefont {O.~G.}\ \bibnamefont
  {Benvenuto}}\ and\ \bibinfo {author} {\bibfnamefont {G.}~\bibnamefont
  {Lugones}},\ }\bibfield  {title} {\bibinfo {title} {{The properties of
  strange stars in the quark mass-density-dependent model}},\ }\href
  {https://doi.org/10.1142/S0218271898000048} {\bibfield  {journal} {\bibinfo
  {journal} {Int. J. Mod. Phys. D}\ }\textbf {\bibinfo {volume} {7}},\ \bibinfo
  {pages} {29} (\bibinfo {year} {1998})}\BibitemShut {NoStop}%
\bibitem [{\citenamefont {Lugones}\ and\ \citenamefont
  {Horvath}(2003)}]{Lugones:2002vd}%
  \BibitemOpen
  \bibfield  {author} {\bibinfo {author} {\bibfnamefont {G.}~\bibnamefont
  {Lugones}}\ and\ \bibinfo {author} {\bibfnamefont {J.~E.}\ \bibnamefont
  {Horvath}},\ }\bibfield  {title} {\bibinfo {title} {{Quark - diquark equation
  of state and compact star structure}},\ }\href
  {https://doi.org/10.1142/S0218271803002755} {\bibfield  {journal} {\bibinfo
  {journal} {Int. J. Mod. Phys. D}\ }\textbf {\bibinfo {volume} {12}},\
  \bibinfo {pages} {495} (\bibinfo {year} {2003})},\ \Eprint
  {https://arxiv.org/abs/astro-ph/0203069} {arXiv:astro-ph/0203069}
  \BibitemShut {NoStop}%
\bibitem [{\citenamefont {Peng}\ \emph
  {et~al.}(2000{\natexlab{a}})\citenamefont {Peng}, \citenamefont {Chiang},
  \citenamefont {Yang}, \citenamefont {Li},\ and\ \citenamefont
  {Liu}}]{Peng:1999gh}%
  \BibitemOpen
  \bibfield  {author} {\bibinfo {author} {\bibfnamefont {G.~X.}\ \bibnamefont
  {Peng}}, \bibinfo {author} {\bibfnamefont {H.~C.}\ \bibnamefont {Chiang}},
  \bibinfo {author} {\bibfnamefont {J.~J.}\ \bibnamefont {Yang}}, \bibinfo
  {author} {\bibfnamefont {L.}~\bibnamefont {Li}},\ and\ \bibinfo {author}
  {\bibfnamefont {B.}~\bibnamefont {Liu}},\ }\bibfield  {title} {\bibinfo
  {title} {{Mass formulas and thermodynamic treatment in the mass density
  dependent model of strange quark matter}},\ }\href
  {https://doi.org/10.1103/PhysRevC.61.015201} {\bibfield  {journal} {\bibinfo
  {journal} {Phys. Rev. C}\ }\textbf {\bibinfo {volume} {61}},\ \bibinfo
  {pages} {015201} (\bibinfo {year} {2000}{\natexlab{a}})},\ \Eprint
  {https://arxiv.org/abs/hep-ph/9911222} {arXiv:hep-ph/9911222} \BibitemShut
  {NoStop}%
\bibitem [{\citenamefont {Wang}(2000)}]{Wang:2000dc}%
  \BibitemOpen
  \bibfield  {author} {\bibinfo {author} {\bibfnamefont {P.}~\bibnamefont
  {Wang}},\ }\bibfield  {title} {\bibinfo {title} {{Strange matter in a
  selfconsistent quark mass density dependent model}},\ }\href
  {https://doi.org/10.1103/PhysRevC.62.015204} {\bibfield  {journal} {\bibinfo
  {journal} {Phys. Rev. C}\ }\textbf {\bibinfo {volume} {62}},\ \bibinfo
  {pages} {015204} (\bibinfo {year} {2000})}\BibitemShut {NoStop}%
\bibitem [{\citenamefont {Peng}\ \emph
  {et~al.}(2000{\natexlab{b}})\citenamefont {Peng}, \citenamefont {Chiang},\
  and\ \citenamefont {Ning}}]{Peng:2000ff}%
  \BibitemOpen
  \bibfield  {author} {\bibinfo {author} {\bibfnamefont {G.~X.}\ \bibnamefont
  {Peng}}, \bibinfo {author} {\bibfnamefont {H.~C.}\ \bibnamefont {Chiang}},\
  and\ \bibinfo {author} {\bibfnamefont {P.~Z.}\ \bibnamefont {Ning}},\
  }\bibfield  {title} {\bibinfo {title} {{Thermodynamics, strange quark matter,
  and strange stars}},\ }\href {https://doi.org/10.1103/PhysRevC.62.025801}
  {\bibfield  {journal} {\bibinfo  {journal} {Phys. Rev. C}\ }\textbf {\bibinfo
  {volume} {62}},\ \bibinfo {pages} {025801} (\bibinfo {year}
  {2000}{\natexlab{b}})},\ \Eprint {https://arxiv.org/abs/hep-ph/0003027}
  {arXiv:hep-ph/0003027} \BibitemShut {NoStop}%
\bibitem [{\citenamefont {Yin}\ and\ \citenamefont {Su}(2008)}]{Yin:2008me}%
  \BibitemOpen
  \bibfield  {author} {\bibinfo {author} {\bibfnamefont {S.-y.}\ \bibnamefont
  {Yin}}\ and\ \bibinfo {author} {\bibfnamefont {R.-K.}\ \bibnamefont {Su}},\
  }\bibfield  {title} {\bibinfo {title} {{Consistent thermodynamic treatment
  for a quark-mass density-dependent model}},\ }\href
  {https://doi.org/10.1103/PhysRevC.77.055204} {\bibfield  {journal} {\bibinfo
  {journal} {Phys. Rev. C}\ }\textbf {\bibinfo {volume} {77}},\ \bibinfo
  {pages} {055204} (\bibinfo {year} {2008})},\ \Eprint
  {https://arxiv.org/abs/0801.2813} {arXiv:0801.2813 [nucl-th]} \BibitemShut
  {NoStop}%
\bibitem [{\citenamefont {Callen}(1985)}]{Callen}%
  \BibitemOpen
  \bibfield  {author} {\bibinfo {author} {\bibfnamefont {H.}~\bibnamefont
  {Callen}},\ }\href@noop {} {\emph {\bibinfo {title} {Thermodynamics and an
  Introduction to Thermostatistics}}}\ (\bibinfo  {publisher} {Wiley},\
  \bibinfo {year} {1985})\BibitemShut {NoStop}%
\bibitem [{\citenamefont {Xia}\ \emph {et~al.}(2014)\citenamefont {Xia},
  \citenamefont {Peng}, \citenamefont {Chen}, \citenamefont {Lu},\ and\
  \citenamefont {Xu}}]{Xia:2014zaa}%
  \BibitemOpen
  \bibfield  {author} {\bibinfo {author} {\bibfnamefont {C.~J.}\ \bibnamefont
  {Xia}}, \bibinfo {author} {\bibfnamefont {G.~X.}\ \bibnamefont {Peng}},
  \bibinfo {author} {\bibfnamefont {S.~W.}\ \bibnamefont {Chen}}, \bibinfo
  {author} {\bibfnamefont {Z.~Y.}\ \bibnamefont {Lu}},\ and\ \bibinfo {author}
  {\bibfnamefont {J.~F.}\ \bibnamefont {Xu}},\ }\bibfield  {title} {\bibinfo
  {title} {{Thermodynamic consistency, quark mass scaling, and properties of
  strange matter}},\ }\href {https://doi.org/10.1103/PhysRevD.89.105027}
  {\bibfield  {journal} {\bibinfo  {journal} {Phys. Rev. D}\ }\textbf {\bibinfo
  {volume} {89}},\ \bibinfo {pages} {105027} (\bibinfo {year} {2014})},\
  \Eprint {https://arxiv.org/abs/1405.3037} {arXiv:1405.3037 [hep-ph]}
  \BibitemShut {NoStop}%
\bibitem [{\citenamefont {Gorenstein}\ and\ \citenamefont
  {Yang}(1995)}]{Gorenstein:1995vm}%
  \BibitemOpen
  \bibfield  {author} {\bibinfo {author} {\bibfnamefont {M.~I.}\ \bibnamefont
  {Gorenstein}}\ and\ \bibinfo {author} {\bibfnamefont {S.-N.}\ \bibnamefont
  {Yang}},\ }\bibfield  {title} {\bibinfo {title} {{Gluon plasma with a medium
  dependent dispersion relation}},\ }\href
  {https://doi.org/10.1103/PhysRevD.52.5206} {\bibfield  {journal} {\bibinfo
  {journal} {Phys. Rev. D}\ }\textbf {\bibinfo {volume} {52}},\ \bibinfo
  {pages} {5206} (\bibinfo {year} {1995})}\BibitemShut {NoStop}%
\bibitem [{\citenamefont {Schertler}\ \emph {et~al.}(1997)\citenamefont
  {Schertler}, \citenamefont {Greiner},\ and\ \citenamefont
  {Thoma}}]{Schertler:1996tq}%
  \BibitemOpen
  \bibfield  {author} {\bibinfo {author} {\bibfnamefont {K.}~\bibnamefont
  {Schertler}}, \bibinfo {author} {\bibfnamefont {C.}~\bibnamefont {Greiner}},\
  and\ \bibinfo {author} {\bibfnamefont {M.~H.}\ \bibnamefont {Thoma}},\
  }\bibfield  {title} {\bibinfo {title} {{Medium effects in strange quark
  matter and strange stars}},\ }\href
  {https://doi.org/10.1016/S0375-9474(97)00014-6} {\bibfield  {journal}
  {\bibinfo  {journal} {Nucl. Phys. A}\ }\textbf {\bibinfo {volume} {616}},\
  \bibinfo {pages} {659} (\bibinfo {year} {1997})},\ \Eprint
  {https://arxiv.org/abs/hep-ph/9611305} {arXiv:hep-ph/9611305} \BibitemShut
  {NoStop}%
\bibitem [{\citenamefont {Wen}\ \emph {et~al.}(2010)\citenamefont {Wen},
  \citenamefont {Li}, \citenamefont {Liang},\ and\ \citenamefont
  {Peng}}]{Wen:2010zz}%
  \BibitemOpen
  \bibfield  {author} {\bibinfo {author} {\bibfnamefont {X.~J.}\ \bibnamefont
  {Wen}}, \bibinfo {author} {\bibfnamefont {J.~Y.}\ \bibnamefont {Li}},
  \bibinfo {author} {\bibfnamefont {J.~Q.}\ \bibnamefont {Liang}},\ and\
  \bibinfo {author} {\bibfnamefont {G.~X.}\ \bibnamefont {Peng}},\ }\bibfield
  {title} {\bibinfo {title} {{Medium effects on the surface tension of
  strangelets in the extended quasiparticle model}},\ }\href
  {https://doi.org/10.1103/PhysRevC.82.025809} {\bibfield  {journal} {\bibinfo
  {journal} {Phys. Rev. C}\ }\textbf {\bibinfo {volume} {82}},\ \bibinfo
  {pages} {025809} (\bibinfo {year} {2010})}\BibitemShut {NoStop}%
\bibitem [{\citenamefont {Shapiro}\ and\ \citenamefont
  {Teukolsky}(1983)}]{Shapiro:1983du}%
  \BibitemOpen
  \bibfield  {author} {\bibinfo {author} {\bibfnamefont {S.~L.}\ \bibnamefont
  {Shapiro}}\ and\ \bibinfo {author} {\bibfnamefont {S.~A.}\ \bibnamefont
  {Teukolsky}},\ }\href@noop {} {\emph {\bibinfo {title} {{Black holes, white
  dwarfs, and neutron stars: The physics of compact objects}}}}\ (\bibinfo
  {publisher} {Wiley},\ \bibinfo {year} {1983})\BibitemShut {NoStop}%
\bibitem [{\citenamefont {Kaltenborn}\ \emph {et~al.}(2017)\citenamefont
  {Kaltenborn}, \citenamefont {Bastian},\ and\ \citenamefont
  {Blaschke}}]{Kaltenborn:2017hus}%
  \BibitemOpen
  \bibfield  {author} {\bibinfo {author} {\bibfnamefont {M.~A.~R.}\
  \bibnamefont {Kaltenborn}}, \bibinfo {author} {\bibfnamefont {N.-U.~F.}\
  \bibnamefont {Bastian}},\ and\ \bibinfo {author} {\bibfnamefont {D.~B.}\
  \bibnamefont {Blaschke}},\ }\bibfield  {title} {\bibinfo {title}
  {{Quark-nuclear hybrid star equation of state with excluded volume
  effects}},\ }\href {https://doi.org/10.1103/PhysRevD.96.056024} {\bibfield
  {journal} {\bibinfo  {journal} {Phys. Rev.}\ }\textbf {\bibinfo {volume}
  {D96}},\ \bibinfo {pages} {056024} (\bibinfo {year} {2017})},\ \Eprint
  {https://arxiv.org/abs/1701.04400} {arXiv:1701.04400 [astro-ph.HE]}
  \BibitemShut {NoStop}%
%%CITATION = ARXIV:1701.04400;%%
\bibitem [{\citenamefont {Wen}\ \emph {et~al.}(2005)\citenamefont {Wen},
  \citenamefont {Zhong}, \citenamefont {Peng}, \citenamefont {Shen},\ and\
  \citenamefont {Ning}}]{Wen:2005uf}%
  \BibitemOpen
  \bibfield  {author} {\bibinfo {author} {\bibfnamefont {X.~J.}\ \bibnamefont
  {Wen}}, \bibinfo {author} {\bibfnamefont {X.~H.}\ \bibnamefont {Zhong}},
  \bibinfo {author} {\bibfnamefont {G.~X.}\ \bibnamefont {Peng}}, \bibinfo
  {author} {\bibfnamefont {P.~N.}\ \bibnamefont {Shen}},\ and\ \bibinfo
  {author} {\bibfnamefont {P.~Z.}\ \bibnamefont {Ning}},\ }\bibfield  {title}
  {\bibinfo {title} {{Thermodynamics with density and temperature dependent
  particle masses and properties of bulk strange quark matter and
  strangelets}},\ }\href {https://doi.org/10.1103/PhysRevC.72.015204}
  {\bibfield  {journal} {\bibinfo  {journal} {Phys. Rev. C}\ }\textbf {\bibinfo
  {volume} {72}},\ \bibinfo {pages} {015204} (\bibinfo {year} {2005})},\
  \Eprint {https://arxiv.org/abs/hep-ph/0506050} {arXiv:hep-ph/0506050}
  \BibitemShut {NoStop}%
\bibitem [{\citenamefont {Pisarski}(1989)}]{Pisarski:1989wb}%
  \BibitemOpen
  \bibfield  {author} {\bibinfo {author} {\bibfnamefont {R.~D.}\ \bibnamefont
  {Pisarski}},\ }\bibfield  {title} {\bibinfo {title} {{Renormalized Fermion
  Propagator in Hot Gauge Theories}},\ }\href
  {https://doi.org/10.1016/0375-9474(89)90620-9} {\bibfield  {journal}
  {\bibinfo  {journal} {Nucl. Phys. A}\ }\textbf {\bibinfo {volume} {498}},\
  \bibinfo {pages} {423C} (\bibinfo {year} {1989})}\BibitemShut {NoStop}%
\bibitem [{\citenamefont {Blaizot}\ and\ \citenamefont
  {Ollitrault}(1993)}]{Blaizot:1993bb}%
  \BibitemOpen
  \bibfield  {author} {\bibinfo {author} {\bibfnamefont {J.-P.}\ \bibnamefont
  {Blaizot}}\ and\ \bibinfo {author} {\bibfnamefont {J.-Y.}\ \bibnamefont
  {Ollitrault}},\ }\bibfield  {title} {\bibinfo {title} {{Collective fermionic
  excitations in systems with a large chemical potential}},\ }\href
  {https://doi.org/10.1103/PhysRevD.48.1390} {\bibfield  {journal} {\bibinfo
  {journal} {Phys. Rev. D}\ }\textbf {\bibinfo {volume} {48}},\ \bibinfo
  {pages} {1390} (\bibinfo {year} {1993})},\ \Eprint
  {https://arxiv.org/abs/hep-th/9303070} {arXiv:hep-th/9303070} \BibitemShut
  {NoStop}%
\bibitem [{\citenamefont {Schertler}\ \emph {et~al.}(1998)\citenamefont
  {Schertler}, \citenamefont {Greiner}, \citenamefont {Sahu},\ and\
  \citenamefont {Thoma}}]{Schertler:1997za}%
  \BibitemOpen
  \bibfield  {author} {\bibinfo {author} {\bibfnamefont {K.}~\bibnamefont
  {Schertler}}, \bibinfo {author} {\bibfnamefont {C.}~\bibnamefont {Greiner}},
  \bibinfo {author} {\bibfnamefont {P.~K.}\ \bibnamefont {Sahu}},\ and\
  \bibinfo {author} {\bibfnamefont {M.~H.}\ \bibnamefont {Thoma}},\ }\bibfield
  {title} {\bibinfo {title} {{The Influence of medium effects on the gross
  structure of hybrid stars}},\ }\href
  {https://doi.org/10.1016/S0375-9474(98)00330-3} {\bibfield  {journal}
  {\bibinfo  {journal} {Nucl. Phys. A}\ }\textbf {\bibinfo {volume} {637}},\
  \bibinfo {pages} {451} (\bibinfo {year} {1998})},\ \Eprint
  {https://arxiv.org/abs/astro-ph/9712165} {arXiv:astro-ph/9712165}
  \BibitemShut {NoStop}%
\bibitem [{\citenamefont {Chu}\ \emph {et~al.}(2021)\citenamefont {Chu},
  \citenamefont {Zhou}, \citenamefont {Jiang}, \citenamefont {Ma},
  \citenamefont {Liu}, \citenamefont {Zhang},\ and\ \citenamefont
  {Li}}]{Chu:2021vvv}%
  \BibitemOpen
  \bibfield  {author} {\bibinfo {author} {\bibfnamefont {P.-C.}\ \bibnamefont
  {Chu}}, \bibinfo {author} {\bibfnamefont {Y.}~\bibnamefont {Zhou}}, \bibinfo
  {author} {\bibfnamefont {Y.-Y.}\ \bibnamefont {Jiang}}, \bibinfo {author}
  {\bibfnamefont {H.-Y.}\ \bibnamefont {Ma}}, \bibinfo {author} {\bibfnamefont
  {H.}~\bibnamefont {Liu}}, \bibinfo {author} {\bibfnamefont {X.-M.}\
  \bibnamefont {Zhang}},\ and\ \bibinfo {author} {\bibfnamefont {X.-H.}\
  \bibnamefont {Li}},\ }\bibfield  {title} {\bibinfo {title} {{Quark star
  matter in heavy quark stars}},\ }\href
  {https://doi.org/10.1140/epjc/s10052-020-08800-3} {\bibfield  {journal}
  {\bibinfo  {journal} {Eur. Phys. J. C}\ }\textbf {\bibinfo {volume} {81}},\
  \bibinfo {pages} {93} (\bibinfo {year} {2021})}\BibitemShut {NoStop}%
\bibitem [{\citenamefont {Chang}\ \emph {et~al.}(2013)\citenamefont {Chang},
  \citenamefont {Chen}, \citenamefont {Peng},\ and\ \citenamefont
  {Xu}}]{Chang:2013wnl}%
  \BibitemOpen
  \bibfield  {author} {\bibinfo {author} {\bibfnamefont {Q.}~\bibnamefont
  {Chang}}, \bibinfo {author} {\bibfnamefont {S.}~\bibnamefont {Chen}},
  \bibinfo {author} {\bibfnamefont {G.}~\bibnamefont {Peng}},\ and\ \bibinfo
  {author} {\bibfnamefont {J.}~\bibnamefont {Xu}},\ }\bibfield  {title}
  {\bibinfo {title} {{Properties of color-flavor locked strange quark matter
  and strange stars in a new quark mass scaling}},\ }\href
  {https://doi.org/10.1007/s11433-013-5160-z} {\bibfield  {journal} {\bibinfo
  {journal} {Sci. China Phys. Mech. Astron.}\ }\textbf {\bibinfo {volume}
  {56}},\ \bibinfo {pages} {1730} (\bibinfo {year} {2013})}\BibitemShut
  {NoStop}%
\bibitem [{\citenamefont {Zyla}\ \emph {et~al.}(2020)\citenamefont {Zyla} \emph
  {et~al.}}]{ParticleDataGroup:2020ssz}%
  \BibitemOpen
  \bibfield  {author} {\bibinfo {author} {\bibfnamefont {P.~A.}\ \bibnamefont
  {Zyla}} \emph {et~al.} (\bibinfo {collaboration} {Particle Data Group}),\
  }\bibfield  {title} {\bibinfo {title} {{Review of Particle Physics}},\ }\href
  {https://doi.org/10.1093/ptep/ptaa104} {\bibfield  {journal} {\bibinfo
  {journal} {PTEP}\ }\textbf {\bibinfo {volume} {2020}},\ \bibinfo {pages}
  {083C01} (\bibinfo {year} {2020})}\BibitemShut {NoStop}%
\bibitem [{\citenamefont {Farhi}\ and\ \citenamefont
  {Jaffe}(1984)}]{Farhi:1984qu}%
  \BibitemOpen
  \bibfield  {author} {\bibinfo {author} {\bibfnamefont {E.}~\bibnamefont
  {Farhi}}\ and\ \bibinfo {author} {\bibfnamefont {R.~L.}\ \bibnamefont
  {Jaffe}},\ }\bibfield  {title} {\bibinfo {title} {{Strange Matter}},\ }\href
  {https://doi.org/10.1103/PhysRevD.30.2379} {\bibfield  {journal} {\bibinfo
  {journal} {Phys. Rev. D}\ }\textbf {\bibinfo {volume} {30}},\ \bibinfo
  {pages} {2379} (\bibinfo {year} {1984})}\BibitemShut {NoStop}%
\end{thebibliography}%
\end{document}